\documentclass[preprint,review,12pt,authoryear]{elsarticle}
\usepackage[T1]{fontenc}
\usepackage{lmodern}
\usepackage[a4paper,left=30mm,right=24mm,top=24mm,bottom=24mm]{geometry}
\usepackage{amsmath,amssymb,graphicx,booktabs,tabularx,array,float,placeins}
\usepackage{microtype}
\usepackage[hidelinks]{hyperref}
\biboptions{round,semicolon,sort}
\setcitestyle{authoryear,round,aysep={,},citesep={;}}
\providecommand{\doi}[1]{\href{https://doi.org/#1}{\nolinkurl{#1}}}
\renewcommand{\arraystretch}{1.08}
\renewcommand{\topfraction}{.93}
\renewcommand{\bottomfraction}{.82}
\renewcommand{\textfraction}{.07}

\newcommand{\dd}{\mathop{}\!\mathrm{d}}
\newcommand{\SI}{Supplementary Information}

\newcommand{\alttext}[1]{\par\noindent{\small\raggedright\linespread{1}\selectfont\textbf{Alt text:} #1\par}}
\newcommand{\APlanetRows}{%
Europa (cold) & 30 & 97 & 0 & 1434 & 1.440 & 177.9 & Liquid \\
Europa (warm) & 5 & 124 & 0 & 1432 & 1.440 & 183.8 & Liquid \\
Enceladus & 27 & 28 & 0 & 197 & 0.128 & 6.3 & Liquid / porous rock \\
Ganymede & 70 & 361 & 427 & 1773 & 1.766 & 1719.8 & HP ice \\
Callisto & 100 & 139 & 6 & 2165 & 1.298 & 318.1 & HP ice / mixed rock \\
Titan (historic) & 50 & 420 & 2 & 2103 & 1.491 & 741.0 & HP ice \\
Titan (slush) & 100 & 0 & 370 & 2105 & 1.456 & 779.3 & No global liquid \\
Triton (conditional) & 150 & 170 & 0 & 1033 & 0.982 & 280.5 & Liquid \\%
}

\newcommand{\ASourceRows}{%
Europa hot & 1 & 170.8 & 3.29 & $2.36\times10^{-6}$ & 0.3898 & 0.005554 \\
Europa hot & 3 & 170.8 & 3.29 & $3.44\times10^{-4}$ & 13.83 & 3.747 \\
Europa hot & 10 & 170.8 & 3.29 & 0.00528 & 19.35 & 10.21 \\
Europa early hot & 1 & 88.6 & 0.53 & $6.16\times10^{-6}$ & 1.274 & 0.03023 \\
Europa early hot & 3 & 88.6 & 0.53 & $2.30\times10^{-4}$ & 7.332 & 3.824 \\
Europa early hot & 10 & 88.6 & 0.53 & 0.00241 & 7.353 & 6.658 \\
Triton (conditional) & 1 & 149.4 & 9.64 & $1.58\times10^{-6}$ & 0.3396 & 0.005814 \\
Triton (conditional) & 3 & 149.4 & 9.64 & $2.20\times10^{-4}$ & 9.924 & 3.618 \\
Triton (conditional) & 10 & 149.4 & 9.64 & 0.00355 & 14.48 & 9.642 \\
Enceladus dry & 1--10 & 42.2 & 9.20 & 0 & 0 & 0 \\%
}

\newcommand{\AResetRows}{%
Europa, 60 km & 0.9982 & 278.8 & 0.250 & 0.250 & 4.529 & 2.585 & 27.39 \\
Europa, 100 km & 1.0004 & 278.8 & 0.250 & 0.250 & 5.000 & 5.000 & 27.39 \\
Early Europa, 100 km & 0.9999 & 281.6 & 0.250 & 0.250 & 5.000 & 4.373 & 14.39 \\
Triton, 60 km & 1.0009 & 274.7 & 0.235 & 0.235 & 4.608 & 2.309 & 24.27 \\%
}

\newcommand{\UCycleRows}{%
Europa, 60 km & 2.11 & 1316.0 & 27.39 & 64.9 & 4.529 & 29.1 \\
Europa, 100 km & 5.86 & 3655.5 & 27.39 & 64.9 & 5.000 & 32.2 \\
Early-hot Europa, 100 km & 2.68 & 583.4 & 14.39 & 33.7 & 5.000 & 32.2 \\
Triton, 60 km & 1.42 & 3855.8 & 24.27 & 56.8 & 4.608 & 29.6 \\%
}
\newcommand{\UFrontRefineRows}{%
Eu\_hot\_g1mm & 128/256/512/1024 & 4.36/4.30/4.30/4.36 & 0 \\
Eu\_hot\_g10mm & 128/256/512/1024 & 4.69/4.31/4.49/4.40 & 0 \\
Tr\_hot\_g10mm & 128/256/512/1024 & 3.13/3.13/2.94/2.96 & 0 \\
S080\_relax & 128/256/512/1024 & 32.83/33.20/33.01/33.01 & 0 \\%
}

\newcommand{\UStructureRows}{%
Eu\_shallow & 3333 & 3333 & -- & 0.3750 & 0.3460 & 0.30 & matched geometry only \\
Eu\_V18\_cold & 3592 & 3431 & 7867 & 0.3460 & 0.3460 & 0.30 & direct liquid \\
Eu\_V18\_warm & 3597 & 3437 & 7806 & 0.3460 & 0.3460 & 0.30 & direct liquid \\
En\_P24 & 2323 & 2323 & -- & 0.3361 & 0.3380 & 0.30 & porous rock / liquid \\
En\_P24\_alt & 2288 & 2288 & -- & 0.3376 & 0.3380 & 0.29 & porous rock / liquid \\
Ga\_cold & 3579 & 3509 & 7203 & 0.3115 & 0.3115 & 0.30 & HP ice \\
Ga\_mid & 3562 & 3507 & 7203 & 0.3115 & 0.3115 & 0.30 & HP ice \\
Ga\_warm & 3570 & 3511 & 7203 & 0.3115 & 0.3115 & 0.30 & HP ice \\
Ca\_cold & 2138 & 2019 & 2737 & 0.3549 & 0.3549 & 0.30 & HP ice / mixed rock \\
Ca\_warm & 2143 & 2028 & 2723 & 0.3549 & 0.3549 & 0.30 & HP ice / mixed rock \\
Ti\_ocean\_cold & 2541 & 2541 & -- & 0.3422 & 0.3438 & 0.30 & HP ice (historic) \\
Ti\_ocean\_warm & 2537 & 2537 & -- & 0.3433 & 0.3438 & 0.30 & HP ice (historic) \\
Ti\_slush & 2474 & 2394 & 2873 & 0.3438 & 0.3438 & -- & no global liquid in model \\
Tr\_liquid & 3400 & 3400 & -- & 0.3119 & -- & 0.30 & conditional liquid \\
Tr\_frozen & 3428 & 3428 & -- & 0.3101 & -- & -- & solid shell \\%
}

\newcommand{\ADiskRows}{%
S037 & 190.8 & 23.16 & 0.0296 / 3.91 & 6609.1 & 44.91 & 0.750 & 1.000 \\
S080 & 222.9 & 36.89 & 0.7277 / 16.89 & 304.5 & 8.49 & 0.326 & 1.000 \\
S084 & 194.8 & 24.64 & 0.8556 / 21.00 & 227.7 & 5.78 & 0.579 & 1.000 \\%
}

\newcommand{\AColdGeometryRows}{%
S037 & 0.750 & 0.202 & 0.093 & 0.596 & 0.750 & 0.828 \\
S080 & 0.326 & 0.236 & 0.075 & 0.676 & 0.324 & 0.601 \\
S084 & 0.579 & 0.207 & 0.046 & 0.611 & 0.382 & 0.734 \\%
}

\ifdefined\VBCBuildLocal
  \pdfmapfile{=lm.map}
  \pdfmapfile{+cm.map}
  \pdfmapfile{+cmextra.map}
  \pdfmapfile{+symbols.map}
  \pdfmapfile{+latxfont.map}
\fi
\journal{Icarus}
\hypersetup{pdftitle={Weak buoyancy and cold lithospheres decouple interior melting from seafloor volcanism on icy ocean worlds},pdfauthor={Nahiz Hestaphin},pdfsubject={Silicate magma transport on icy ocean worlds}}
\begin{document}
\begin{frontmatter}
\title{Weak buoyancy and cold lithospheres decouple interior melting from seafloor volcanism on icy ocean worlds}
\author[cam,man]{Nahiz Hestaphin\corref{cor1}}
\ead{zh389@cantab.ac.uk}
\cortext[cor1]{Corresponding author.}
\affiliation[cam]{organization={Yusuf Hamied Department of Chemistry, University of Cambridge},
  addressline={Lensfield Road}, city={Cambridge}, postcode={CB2~1EW}, country={United Kingdom}}
\affiliation[man]{organization={Department of Physics and Astronomy, The University of Manchester},
  addressline={Schuster Building, Oxford Road}, city={Manchester}, postcode={M13~9PL}, country={United Kingdom}}
\begin{abstract}
On icy ocean worlds, deep silicate melting can generate sufficient pressure to open fractures while magma freezes before reaching the ocean. We identify a coupled transport regime by linking finite melt segregation and nonlinear rock relaxation to fracture flow, crystallisation and the ice-loaded ocean boundary. In a hot Europa endmember, increasing grain size from 1 to 10 mm raises relaxed excess pressure from 5.55 kPa to 10.2 MPa, overcoming the inherited-fracture opening threshold despite weak buoyancy. For the same 100 m strike and 0.5 m aperture, the cold first-fill rate required by heat balance increases from 0.17--0.69 m$^3$ s$^{-1}$ over terrestrial crustal distances of 1.1--4.5 km to 15--29 m$^3$ s$^{-1}$ across 89--171 km Europa paths, far above the self-supplied flow. Higher ice--ocean pressure further reduces the enthalpy margin of near-solidus magma in the adopted basalt phase law. During approximately 5 kyr of dormancy, the collector pressure returns to its relaxed value, while upper-path walls cool to 275--282 K and the magma path loses liquid connectivity. In two focused Europa histories with mechanically supported pathways, delivery of 0.005 km$^3$ of magma over 0.1 Myr requires 14--27 km$^3$ of sacrificed magma-equivalent preparation material. Weak gravity, extensive cold rock and a pressurised ocean thus make seafloor access a history-dependent process governed by thermal connectivity and its renewal, decoupling recurrent volcanism from melt production at depth.
\end{abstract}
\end{frontmatter}

\section{Introduction}
\label{sec:intro}
An ocean separated from a molten region by a cold rocky lithosphere can remain chemically and thermally connected to its interior through aqueous circulation while receiving little silicate magma. This distinction is central to the evolution of icy ocean worlds. Models of tidal and radiogenic heating allow silicate melting at some epochs of rocky-interior evolution \citep{behounkova2021,trinh2023}; hydrothermal circulation provides a separate pathway for ocean--rock exchange \citep{choblet2017,sekine2015,fisher2024}. The extent to which melt itself reaches the ocean determines a different component of this exchange: the emplacement of fresh volcanic rock and heat at the seafloor.

The transport problem combines physical conditions that occur together on icy worlds. Relative to Earth's active oceanic volcanic systems, the low gravity of many satellites weakens buoyant segregation and fracture ascent; deep melt must traverse a thick cold lithosphere to a water boundary loaded by an ocean and ice shell. \citet{behounkova2021} establish the thermal conditions for silicate melt production in Europa's mantle. \citet{bland2022} delimit demanding conditions for buoyant fracture propagation through Europa's lithosphere, while \citet{green2025} find a restricted magmatic driving force and low intergranular supply in present-day Europa models. Recent estimates also indicate weak tectonic stressing of the rocky seafloor \citep{byrne2026}. Taken together, these studies motivate a question about the transition from a pressure-generating interior to a recurrently accessible ocean boundary. Can rheology compensate for weak buoyancy? Once fracture pressure is attainable, can the same finite source fill and heat a continuous path? Can that path preserve its ability to transmit magma between geologically separated eruptive opportunities?

\looseness=-1
Here we couple finite melt segregation, nonlinear rock accommodation, fracture opening, pressure-dependent magma phase change and a liquid-water receiving boundary. The same absolute pressure, gravity, temperature and melt inventory determine the source and the transport path for a given planetary structure. Matched physical controls then isolate the contributions of gravity, transport length and ice--ocean load. Four dimensionless coordinates and a time-dependent material and enthalpy budget distinguish mechanical access, liquid connectivity and actual delivery. This makes it possible to examine whether interior melt generation and episodic seafloor volcanism can evolve on different clocks.

\section{Planetary interior structures and thermal states}
\label{sec:u_structure}
\subsection{Reference structures and ocean--rock interfaces}
\label{sec:columns}
We construct layered spherical reference interiors using measured radius and gravitational parameter \citep{jpl2026} and representative ice, liquid, rock and inner-core configurations \citep{vanceinterior2018,park2024,gaeman2012}. For radial coordinate $r$, enclosed mass $M(r)$, density $\rho(r)$ and absolute pressure $P(r)$, the structural fields satisfy
\begin{align}
\frac{\dd M}{\dd r}&=4\pi r^2\rho, &
g(r)&=\frac{GM(r)}{r^2},\nonumber\\
\frac{\dd P}{\dd r}&=-\rho g(r).
\label{eq:u_hydro}
\end{align}
Mass and polar moment, $C_I=(8\pi/3)\int\rho r^4\dd r$, constrain the bulk layer densities independently where a moment is available. Layer pressures are continuous across interfaces. Cold-liquid water properties follow IAPWS-95 and its melting boundaries \citep{iapws2018,iapwsmelt}; compressed ice phases use phase-specific densities \citep{bezacier2014,journaux2020}. The detailed mass, moment and phase audit is in \SI\ (SI) Section~\ref{si:u_structure}.

\begin{table}[!htbp]
\caption{Reference planetary structures and direct-contact transport domains. All columns close the measured mass with the adopted layering. $I,D,J$ denote ordinary ice, liquid ocean and high-pressure (HP) ice or slush thickness; $R_r$ is rock radius, $g_b$ and $P_b$ are gravity and absolute pressure at the rocky boundary. The historical Titan geometry and the conditional liquid-ocean Triton are specified endmembers. Independent mass, moment and density tests are reported in SI Section~\ref{si:u_structure}.}
\label{tab:u_worlds}
\centering\small\setlength{\tabcolsep}{3pt}
\begin{tabularx}{\linewidth}{@{}>{\raggedright\arraybackslash}Xrrrrrr>{\raggedright\arraybackslash}X@{}}
\toprule
Body & \shortstack[r]{$I$\\(km)}&\shortstack[r]{$D$\\(km)}&\shortstack[r]{$J$\\(km)}&\shortstack[r]{$R_r$\\(km)}&\shortstack[r]{$g_b$\\(m s$^{-2}$)}&\shortstack[r]{$P_b$\\(MPa)}&Interface\\
\midrule
\APlanetRows
\bottomrule
\end{tabularx}
\end{table}

The adopted structures of six icy satellites span substantially different ocean--rock interfaces (Fig.~\ref{fig:structure}a and Table~\ref{tab:u_worlds}). Within the moment-compatible Europa columns, rock-interface gravity is about 1.440 m s$^{-2}$ and absolute pressure is 178--184 MPa. Enceladus instead gives 0.128 m s$^{-2}$ and 6.33 MPa; a conditional liquid-ocean Triton gives 0.982 m s$^{-2}$ and 281 MPa. The selected Ganymede and Callisto columns contain basal high-pressure ice. Titan includes the historical liquid-interior model with a basal high-pressure ice layer and a slush-rich alternative motivated by the tidal-dissipation interpretation of \citet{petricca2025}. Interfaces with intervening solid ice constitute a different transport boundary from direct liquid--rock contact. We examine silicate supply to liquid water in the reference Europa, Enceladus and conditional Triton structures.

\subsection{Rocky thermal evolution and source eligibility}
\label{sec:rock}
We evolve each specified thermal endmember using spherical conductive and latent-heat transport in its rocky layer:
\begin{equation}
\partial_t[\rho_r(c_pT+L_m\phi_m)]=
\frac{1}{r^2}\frac{\partial}{\partial r}
\left(r^2k_r\frac{\partial T}{\partial r}\right)+\mathcal Q_r .
\label{eq:u_rock}
\end{equation}
Here $\rho_r$ is host density, $c_p$ specific heat, $L_m$ melting latent heat, $k_r$ conductivity and $\mathcal Q_r$ the allocated rock heating per unit volume (W m$^{-3}$). The rocky surface is held at its reference boundary temperature; the inner boundary is insulating. The finite-volume implementation and heat balance use shell volumes and face areas from the same spherical structure. The dry lherzolite melting closure follows \citet{katz2003}, with the numerical solidus
\begin{equation}
T_{s,m}=(273.15+1085.7+132.9P_G-5.1P_G^2)\ {\rm K},
\label{eq:drysolidus}
\end{equation}
where $P_G=P/(1\,{\rm GPa})$. The connected segregation interval is restricted to $\phi_m=0.005$--0.225 and $P_G\leq8$. We extract this depth interval from the dated profiles, and define a common fixed source for the long histories by intersecting its depth ranges at 0 and 0.1 Myr. Its top provides the collector depth and initial rock-path length $L$; the intersection fixes the available melt-bearing thickness. Thermal properties and supply curves evolve between the dated fields. The accessible initial stock $S_0$ is 2 per cent of the smaller connected melt inventory within the specified collection area. Subsequent extraction draws on that stock.

\begin{figure}[!htbp]
\centering\includegraphics[width=\textwidth]{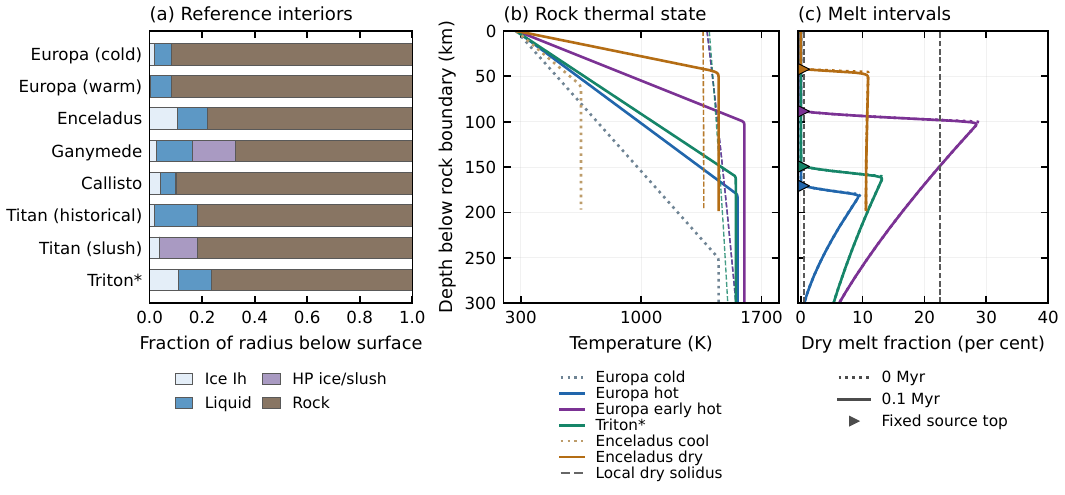}
\caption{Planetary structures and thermal domains. (a) Adopted ordinary ice, liquid, HP ice/slush and rock thicknesses, normalised by each body's radius, for the reference columns in Table~\ref{tab:u_worlds}. (b) Dated rock temperature profiles (solid or dotted) and corresponding pressure-dependent dry solidus (dashed) at 0.1 Myr. Depth is below the rocky boundary, not the outer ice surface. (c) Melt fraction for four thermal endmembers at 0 and 0.1 Myr; vertical lines delimit the connected melt-fraction range and triangles identify the tops of the fixed source intervals at the lower melt-fraction criterion. Enceladus dry denotes a heated porous-core sensitivity; Triton* assumes a liquid ocean. These inherited thermal states test finite-time evolution rather than reconstructing full planetary thermal histories.}
\label{fig:structure}
\alttext{Stacked radius bars distinguish direct liquid--rock contact from interiors containing basal ice or slush. Temperature and melt-fraction profiles show connected melting beginning at different depths in the heated Europa, Enceladus and conditional Triton interiors.}
\end{figure}

The initial rocky thermal profiles have a linear lid over a hot interior. The Europa cold, hot and early-hot reference states start with interior temperatures 1450, 1560 and 1600 K and lids 250, 180 and 100 km, respectively, under the heating allocations detailed in SI Section~\ref{si:u_structure}. The conditional hot Triton endmember starts at 1550 K with a 160 km lid. The connected melting intervals at the two dates determine actual fixed collector depths of 170.8 km (hot Europa), 88.6 km (early-hot Europa) and 149.4 km (hot Triton), distinct from their stipulated initial thermal-lid thicknesses (Fig.~\ref{fig:structure}b,c). The cool Enceladus endmember lacks connected dry silicate melt. Heating the same low-density porous rock (2323 kg m$^{-3}$) to dry melting conditions produces a source-interval sensitivity at 42.2 km but gives no positive basalt buoyancy for a reference 2800 kg m$^{-3}$ melt. The hot Europa and conditional Triton cases instead support segregation in dense host rock. Thus the choice of thermal and structural domain precedes a calculation of fracture transport.

\section{Coupled magma transport model}
\label{sec:u_chain}
\subsection{Melt segregation, finite stock and collector rheology}
\label{sec:source}
For depth $z$ measured downward from the top of the connected source, positive upward relative flux $q$ (m s$^{-1}$), and matrix compaction traction $c=\zeta\partial_z q$ (Pa), the local segregation law is \citep{mckenzie1984,miller2014,havlin2013}
\begin{align}
q&=\frac{k_\phi}{\mu_l}\left[(\rho_r-\rho_l)g+\frac{\partial c}{\partial z}\right],
\qquad k_\phi=\frac{d^2\phi_m^3}{100},
\label{eq:darcy}\\
\frac{\partial q}{\partial z}&=2[a_d\sigma_m+a_s|\sigma_m|^{n-1}\sigma_m],
\quad \sigma_m=\phi_m c,\quad n=3.5 .
\label{eq:u_comp}
\end{align}
Here $k_\phi$ is permeability (m$^2$), $d$ grain size, $\mu_l$ liquid dynamic viscosity and $\rho_l$ liquid density. Dry-olivine diffusion and dislocation creep act in parallel \citep{hirth2004,schulz2020}; the diffusion coefficient $a_d\propto d^{-3}$ and both coefficients incorporate thermally activated creep, pressure dependence and melt weakening. SI Section~\ref{si:u_source} specifies their material coefficients and numerical stress inversion. Top traction is $c(0)=p_c$, while basal relative flux vanishes. The volumetric supply $Q_s(p_c)=A_cq(0)$ uses the catchment area $A_c$. Its unloaded rate is $Q_0=Q_s(0)$, and its stopping pressure $P_*$ satisfies $Q_s(P_*)=0$.

Melt entering a collector of radius $R_c$ partitions between elastic storage, viscous accommodation and fracture flow. The accommodation rate $Q_a$ is obtained by inverting the same combined creep law around the cavity. Radial strain rate is $\dot\epsilon_{\rm II}=\sqrt{3}Q_a/(4\pi r^3)$ and integration of the angularly averaged stress gives
\begin{equation}
p_c(Q_a)=2\sqrt{3}\int_{R_c}^{R_{\rm out}}
\langle\sigma_{\rm II}(r,\theta)\rangle_\theta\,\frac{\dd r}{r}.
\label{eq:cavity}
\end{equation}
The finite elastic source, with compliance $C_e=V_c[1/K_l+3/(4G_r)]$ in m$^3$ Pa$^{-1}$, obeys
\begin{align}
C_e\dot p_c&=Q_s(p_c,T,\phi_m,S)-Q_a(p_c,T)-Q_f,\nonumber\\
\dot S&=-Q_s,\qquad S\geq0,\quad p_c\geq0,
\label{eq:u_collect}
\end{align}
where $V_c$ is cavity volume, $K_l=10$ GPa liquid bulk modulus, $G_r=30$ GPa host shear modulus and $Q_f$ route flow. Pressure $p_c$ is excess over the local lithostatic reference. In the sealed collector the relaxed pressure $p_\infty$ satisfies $Q_s(p_\infty)=Q_a(p_\infty)$ with the frozen-date source properties. Sustaining a pressure $p_{\rm req}$ requires both $p_{\rm req}<P_*$ and net charging supply $Q_s(p_{\rm req})>Q_a(p_{\rm req})$. Discharge consumes the difference and stock decreases; no extra melt production is credited during extraction. This balance couples grain-scale flow and nonlinear rock accommodation before invoking a fracture.

\subsection{Fracture opening, hydraulic resistance and phase change}
\label{sec:path}
The mechanical screen for an elongated slot of strike $W$, mean aperture $b$ and inherited seed half-length $a$ uses
\begin{equation}
p_{\rm open}=\left[\max\left(p_{\rm tensile},
\frac{K_{Ic}}{\sqrt{\pi a}},\frac{E'b}{\pi W/2}\right)
-\sigma_T\right]_+ .
\label{eq:u_open}
\end{equation}
Here $K_{Ic}$ is fracture toughness (Pa m$^{1/2}$), $E'$ the plane-strain elastic modulus, $\sigma_T$ the resolved background tension and $p_{\rm tensile}$ the intact-rock tensile-strength term. We write $[y]_+=\max(0,y)$ and $[y]_a^b=\min(b,\max(a,y))$. The tensile-strength term is omitted for an inherited liquid-accessible trace. A held slit represents a separate, mechanically supported transport geometry. The reference elastic parameters are $E'=70/(1-0.25^2)$ GPa, $K_{Ic}=1$ MPa m$^{1/2}$ and nominal dry strength 2 MPa, with alternate circular-flaw, stress and support tests in SI Sections~\ref{si:u_fracture} and \ref{si:u_path}. As a terrestrial strike-scale comparison, mapped fissure segments in the selected bathymetrically supported Axial H3 outline sum to about 296 m \citep{axialoutlines,lesaout2020}. The deep slots use specified strike widths and apertures.

For current wetted length $x$ and coordinate $s$ directed upward from the collector, the head budget is
\begin{align}
p_c+B(x)&=p_{\rm rec}+\Delta p_{\rm visc}
 +\Delta p_{\rm kin}+\Delta p_{\rm residual}, \label{eq:u_head}\\
B(x)&=\int_0^x(\rho_r-\rho_l)g\,\dd s, \nonumber\\
\Delta p_{\rm visc}&=Q_f\int_0^x\frac{12\mu_{\rm eff}}{Wb^3}\,\dd s,\nonumber\\
\Delta p_{\rm kin}&=\frac{\rho_l}{2}\left(\frac{Q_f}{Wb}\right)^2 .
\nonumber
\end{align}
The term $B$ is integrated buoyancy in Pa, $p_{\rm rec}$ is gauge head at the receiver, and $\Delta p_{\rm residual}$ represents additional front or support work. The first three loss terms convert actual volumetric flow into pressure demand; buoyancy and collector elasticity provide driving head. The absolute pressure $P$ sets local phase state and material properties, whereas pressure in equations (\ref{eq:u_open})--(\ref{eq:u_head}) measures the mechanical demand relative to the ambient stress.

The reference transported basalt endmember uses a pressure-shifted solidus $T_s$ distinct from the dry mantle solidus in equation (\ref{eq:drysolidus}):
\begin{align}
T_s(P)&=1325\ {\rm K}+(50\ {\rm K/GPa})P,\nonumber\\
f_l&=\left[\frac{T-T_s(P)}{170\ {\rm K}}\right]_0^1,\qquad
h_l=c_pT+L_ff_l .
\label{eq:u_h}
\end{align}
The specific enthalpy $h_l$ (J kg$^{-1}$) includes latent heat $L_f=4\times10^5$ J kg$^{-1}$. Crystal-bearing viscosity is described using a bounded thermally activated liquid law and packing factor \citep{giordano2008,costa2009,krieger1959}:
\begin{align}
T^\dagger&=\max(T,500\ {\rm K}),\nonumber\\
X_\mu&=\frac{E_\mu}{R_g}
\left(\frac{1}{T^\dagger}-\frac{1}{T_{\rm ref}}\right)
+\frac{PV_\mu}{R_gT^\dagger},\nonumber\\
\mu_{\rm eff}&=\mu_0\exp([X_\mu]_{-20}^{50})
\left[\max\left(10^{-4},1-\frac{1-f_l}{0.6}\right)\right]^{-1.5}.
\label{eq:basalt_visc}
\end{align}
Here $\mu_0=20$ Pa s, $E_\mu=200$ kJ mol$^{-1}$, $V_\mu=5\times10^{-6}$ m$^3$ mol$^{-1}$, $T_{\rm ref}=1500$ K and $R_g$ is the molar gas constant. The liquid path disconnects when $f_l$ falls below 0.4. The corresponding connectivity temperature is $T_{\rm connect}(P)=T_s(P)+0.4(170\ {\rm K})$. Transient heat uptake by each newly wetted wall has conductance $h_c=2k_r/b$ and a cumulative energy per area $E_A(t)$ from the Robin half-space solution given in SI Section~\ref{si:u_fracture}. These relations link local crystallisation and viscosity to progressively cooled hydraulic resistance.

\subsection{Paired transport margins, time integration and verification}
\label{sec:margins}
To compare transport domains, we use four dimensionless margins for a given source, path and receiver target:
\begin{align}
\Pi_P&=\frac{p_\infty}{p_{\rm open}}, &
\Pi_H&=\frac{H_{\rm available}}{H_{\rm required}},\nonumber\\
\Pi_V&=\frac{V_{\rm available}}{V_{\rm required}}, &
\Pi_T&=\frac{t_{\rm dorm}}{t_{\rm cool}} .
\label{eq:state}
\end{align}
The pressure ratio applies to positive opening thresholds; an accessible zero-threshold path enters the hydraulic calculation directly. The other two delivery ratios use the same $W$, $b$, $L$, bed target $V_*$ and time window $t_w$. To establish a favourable necessary envelope for cold transport, we calculate the fully wetted isothermal loaded flux $Q_h$ as the net source supply $Q_s(p_\ell)-Q_a(p_\ell)$ at the corresponding loading pressure $p_\ell$:
\begin{align}
p_\ell=\max\bigg(&0,p_{\rm open},p_{\rm rec}-B(L)
+\frac{12 Q_h}{Wb^3}\int_0^L\mu_{\rm eff}\,\dd s\nonumber\\
&+\frac{\rho_l}{2}(Q_h/Wb)^2\bigg).
\label{eq:loaded}
\end{align}
The viscosity here is evaluated at inlet temperature and local pressure. Charge time and extracted volume $t_{\rm ch},V_{\rm ch}$ are evaluated by integrating equation (\ref{eq:u_collect}) up to the opening threshold with the dated source curve. For positive loaded flow,
\begin{align}
t_{\rm tr}&=(WbL+V_*)/Q_h,\nonumber\\
V_{\rm required}&=Q_s(p_\ell)t_{\rm tr},\nonumber\\
V_{\rm available}&=\min\{[S_0-V_{\rm ch}]_+,\nonumber\\
&\hspace{2.5em}Q_s(p_\ell)[t_w-t_{\rm ch}]_+\}.
\label{eq:inventory}
\end{align}
The wall exposure at cumulative bed payload $v$ is $t(z';v)=(Wbz'+v)/Q_h$, where $z'$ is distance back from the outlet. Required and available heat are
\begin{align}
H_{\rm required}(v)&=2W\int_0^L E_A(t(z';v))\,\dd z',\nonumber\\
H_{\rm available}(v)&=\rho_l(WbL+v)[h_{\rm in}-h_{\min}]_+,
\label{eq:heat}
\end{align}
where $h_{\min}=\min_{z'}h_l(T_{\rm connect},P)$ is the favourable minimum connected enthalpy along the path. We use the smaller heat ratio at first fill ($v=0$) and at the specified payload ($v=V_*$). Thus $\Pi_P>1$, $\Pi_H\geq1$ and $\Pi_V\geq1$ are optimistic necessary conditions for corresponding positive-threshold, constant-rate cold-path delivery. When a positive loaded rate is unavailable, the latter two ratios are undefined. The full charge and heat integrals, separate imposed-pulse controls and paid-contact margins are derived in SI Section~\ref{si:state}. The memory ratio $\Pi_T$ compares rest duration with the finite-strike thermal diffusion time (Section~\ref{sec:clocks}).

The time-dependent model advances the actual wet length with elastic pressure and stock updated implicitly over each accepted interval. Newly wetted cells receive advective enthalpy, exchange heat through a wall-normal diffusion grid and account for reversible pressure work. The same local phase law determines connectivity and hydraulic resistance. The calculation localises loss of upstream connectivity and the full-length arrival event. The mechanical support and recipient head evolve consistently with accepted flow; the receiving model updates the deposited material, cooling water and elevation-dependent back pressure from the actual volumetric and enthalpy flux. During rest, pressure recovery, cooling of host and plugs, and receiver evolution proceed on the same clock. For a held route, the required edge toughness and compressive support are screened separately from heat balance and paid contact (SI Section~\ref{si:u_path}).

The transferred-volume identity is
\begin{align}
V_x+V_{c0}&=C_ep_c+V_a+V_{\rm wet}+V_{\rm plug}
+V_p+V_b,\nonumber\\
V_b&=V_r+V_{\rm other}.
\label{eq:ledger}
\end{align}
Here $V_x$ is extracted magma and $V_{c0}$ the initially mobile volume including elastic storage and any seed material. The destinations are stored elastic volume, accommodated magma $V_a$, present wet-path volume, earlier frozen plugs, preparation debit $V_p$, and cumulative bed arrival $V_b$ (retained $V_r$ plus other receiver-phase material). The preparation debit is the sacrificed magma-equivalent resource needed for heating: it is distinct from bed emission. The enthalpy budget includes source, magma, walls, prior plugs, preparation residual enthalpy and receiving-water heat; the mechanical ledger accounts for collector and gravitational work, dissipation and receiver work. Detailed interfaces and ledgers appear in SI Sections~\ref{si:u_path} and \ref{si:recurrence}.

We verify the common model against homogeneous diffusion-only compaction, Newtonian cavity and pipe limits, and analytic transient wall heat solutions. Independent thermal, source and fracture reconstructions test alternative geometries and material laws (SI Section~\ref{si:reference_controls}); a published Europa dike trajectory data set gives an external comparison \citep{green2025,green2024data}. The 136 primary, 35 mapped and 40 refinement coupled calculations close relative material and thermal budgets within $6\times10^{-13}$ and $2.0\times10^{-8}$, respectively, with source--path and receiving heat transfers within $3\times10^{-15}$ and integrated pressure work within $5\times10^{-12}$. Spatial and time refinement are evaluated separately using endpoint and flow/temperature comparisons, rather than treating discrete conservation as evidence of solution convergence (SI Section~\ref{si:u_numerics}).

\section{Mechanical and thermal constraints on magma delivery}
\label{sec:constraints}
\subsection{Weak buoyancy and rheological reversal of opening}
\label{sec:u_pressure}
\begin{figure}[!htbp]
\centering\includegraphics[width=\textwidth]{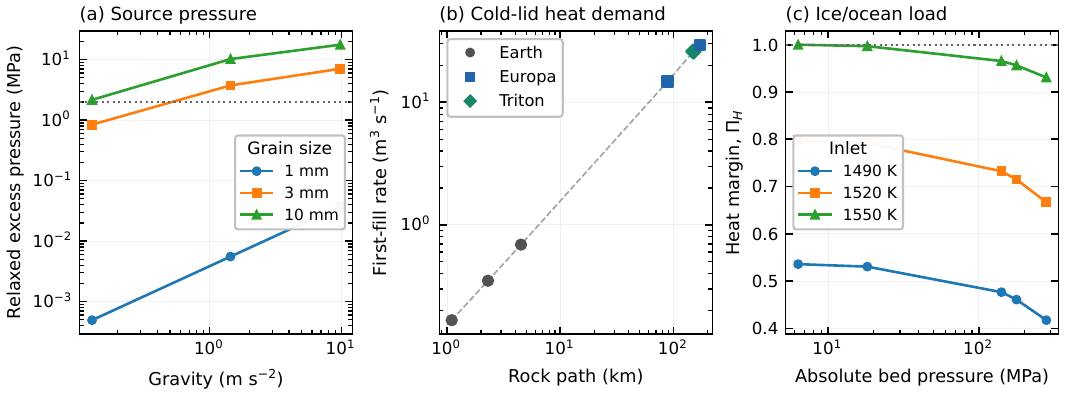}
\caption{Physical drivers of icy-world magma transport. (a) Relaxed collector pressure for the same hot Europa source with varied mechanical gravity and grain size; dotted line is the nominal 2 MPa intact-rock tensile-strength scale. (b) Matched first-fill heat-balance rate for a 100 m by 0.5 m cold fracture with common linear host thermal profile at terrestrial and icy-world path lengths. The connecting line guides comparison between specified lengths; Triton assumes a liquid ocean. (c) Pressure-only comparison of the enthalpy margin for the same 171 km Europa path, imposed packet and receiver target; the pressure field is offset with the absolute floor pressure $P_b$, shifting the reference basalt phase interval.}
\label{fig:drivers}
\alttext{Three plots show higher relaxed collector pressures with increasing gravity and grain size, rising first-fill flow requirements with increasing rock-path length, and decreasing heat margins as the ice--ocean pressure increases. The pressure effect is strongest for cooler magma.}
\end{figure}

The relaxed pressure in the range of connected sources reflects both the stopping pressure set by porous segregation and the competition with collector accommodation (Table~\ref{tab:u_source}). In matched mechanical controls using the same hot Europa material, changing only gravity from 9.81 to 1.440 m s$^{-2}$ reduces relaxed pressure from 37.8 to 5.55 kPa at 1 mm grain size (Fig.~\ref{fig:drivers}a). The stopping pressure decreases from 2.65 to 0.390 MPa. Weak gravity also changes detached fracture ascent: at fixed toughness and density contrast the finite-batch buoyant-fracture volume threshold scales as $g^{-5/3}$ \citep{davis2020,davis2023} (SI Section~\ref{si:reference_controls}).

\begin{table}[!htbp]
\caption{Finite source states from the fixed connected intervals of the dated rock profiles. Grain size $d$, collector depth $L$, accessible inventory $S_0$, unloaded segregation rate $Q_0$, stopping pressure $P_*$ and relaxed pressure $p_\infty$ refer to a 3 km catchment and 1 km collector. The final row combines the three grain sizes (1, 3 and 10 mm) with identical zero buoyant supply in the hot dry Enceladus sensitivity. It retains its non-zero inherited melt stock.}
\label{tab:u_source}
\centering\small
\begin{tabular}{@{}lrrrrrr@{}}
\toprule
Thermal case & $d$ (mm)&$L$ (km)&$S_0$ (km$^3$)&$Q_0$ (m$^3$ s$^{-1}$)&$P_*$ (MPa)&$p_\infty$ (MPa)\\
\midrule
\ASourceRows
\bottomrule
\end{tabular}
\end{table}

Grain size changes both permeability and diffusion creep. At 10 mm, the same hot Europa source develops 10.2 MPa relaxed pressure and $5.28\times10^{-3}$ m$^3$ s$^{-1}$ unloaded flow, compared with $2.36\times10^{-6}$ m$^3$ s$^{-1}$ at 1 mm. At 0.2 MPa the fine-grained collector's accommodation exceeds its loaded inflow by a factor of about 73; the stopping pressure itself lies below the nominal 2 MPa intact-rock strength. Across the tested rheology and geometry endmembers the hot Europa relaxed pressure spans 3.7 Pa to 19.35 MPa (SI Section~\ref{si:u_source}). Coarser grains and nonlinear creep can therefore move the same weak-gravity source from accommodation-dominated behaviour to sufficient fracture pressure. The opening threshold also depends on inherited seed, aperture and stress conditions; a pressure reversal defines mechanical eligibility, not the ensuing transmitted flow.

\subsection{Cold-lithosphere heat demand and the ice--ocean load}
\label{sec:u_freezing}
The magnitude of the thermal constraint is exposed by holding fracture dimensions fixed across rock-path lengths. With 100 m strike, 0.5 m aperture and a common linear host thermal profile, an optimistic first-fill heat balance requires 0.17--0.69 m$^3$ s$^{-1}$ over 1.1--4.5 km, the shallow source-depth scale independently inferred at the active terrestrial Axial Seamount \citep{yang2024}. For the illustrative hot Europa path lengths of 89--171 km the same requirement is 15--29 m$^3$ s$^{-1}$ (Fig.~\ref{fig:drivers}b). For the conditional 149 km Triton path it is about 26 m$^3$ s$^{-1}$. These are matched calculated heat thresholds. For long paths, the heat sink of the newly wetted cold rock consumes the advected sensible and latent enthalpy while the fracture fills. The available porous influx can fall far below the required flow even when sufficient opening pressure is generated.

The ocean and ice load changes magma phase state throughout the path as well as imposing a cold receiver. We isolate absolute pressure at fixed Europa gravity, 171 km path, rock temperature, aperture and imposed source input, varying the entire pressure field consistently with the boundary stress. At 1490 K, increasing the bed pressure from 18.2 to 177.9 MPa lowers the inlet liquid fraction from 0.730 to 0.683 and raises its viscosity from 76 to 102 Pa s in the adopted basalt crystallisation law. At the conditional Triton load of 280.5 MPa the values are 0.653 and 126 Pa s. The heat margin declines from 0.531 to 0.461 to 0.417 (Fig.~\ref{fig:drivers}c). Removing the 36.7 MPa ice-roof load alone in this fixed-column Europa experiment raises the margin to 0.477. These pressure-dependent effects are weaker for hotter liquid and compound the long-lid heat demand, particularly near crystallisation. The reference phase and viscosity sensitivity ranges are documented in SI Section~\ref{si:u_fracture}. The increased phase-change requirement and the associated viscous resistance act in addition to the weak excess-pressure driving force.

\subsection{Joint pressure--heat--inventory constraints}
\label{sec:joint}
\begin{figure}[!htbp]
\centering\includegraphics[width=\textwidth]{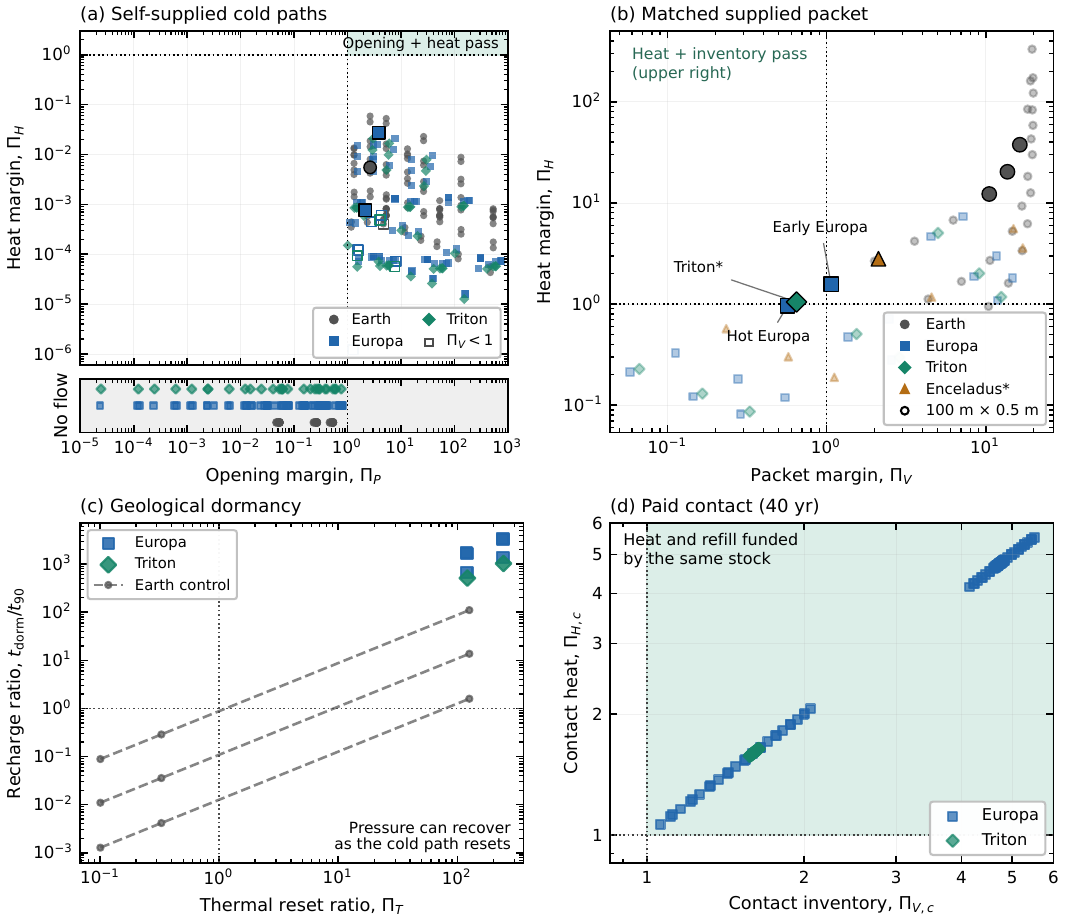}
\caption{Unified transport state comparisons. (a) Self-supplied cold transport for 0.25 million m$^3$ at the bed in 5 kyr, 0.5 MPa gauge receiver and inherited 1 km seed, with varied widths and apertures. Filled markers satisfy $\Pi_V\geq1$; open squares lack inventory/time. The separated grey strip denotes no positive loaded flow (heat and volume margins undefined). Black-edged points indicate the illustrative hot Europa 10 mm, its 100 km focused variant and shallow terrestrial 10 mm cases at $W=1000$ m, $b=0.1$ m. (b) Imposed 5 million m$^3$ in 46 h at 1550 K through held slots, with 0.25 million m$^3$ target; filled black-edged points represent $W=100$ m, $b=0.5$ m. Enceladus* is the dense-host thermal control. (c) Recurrent thermal reset and pressure recharge using the finite-strike cooling scale ($W=100$ m) and sealed-source $t_{90}$. Earth analogs have three grain sizes and specified 4, 13 and 5,000 yr intervals. (d) Margins for 40 yr of paid heat contact and refill in the recurrent histories, using the same finite stock for both requirements. Panels compare distinct experimental designs. Dotted lines indicate unity, a necessary boundary for the relevant margins, not a sufficient condition for dynamic transit. Solid-ice interfaces lie outside the direct-water receiver domain.}
\label{fig:state}
\alttext{Four panels compare transport margins. Cold self-supplied routes remain heat-limited despite sufficient opening pressure. A large imposed pulse approaches or crosses heat and inventory thresholds at long rock-path lengths. Long icy-world rest intervals exceed both recharge and cooling times; the specified renewed-contact cases satisfy the paired heat and inventory margins.}
\end{figure}

The first two controls establish why pressure and temperature alone are insufficient descriptions of the transport regime. We now ask whether actual source and pathway budgets can satisfy the joint margins of equation (\ref{eq:state}). For a common inherited fracture with 1 km seed half-length, 1 km strike, 0.1 m aperture and zero added tensile stress, $p_{\rm open}=4.75$ MPa. For the hot profiles with a 0.5 MPa receiver gauge head, hot Europa moves from $\Pi_P=0.00117$ (1 mm) to 2.15 (10 mm). Its 10 mm loaded rate is $3.80\times10^{-3}$ m$^3$ s$^{-1}$ with $\Pi_V=34.4$ but $\Pi_H=7.62\times10^{-4}$. Focusing a 100 km catchment into a 3 km collector raises the rate to 4.42 m$^3$ s$^{-1}$, with $\Pi_P=3.84$ and $\Pi_H=0.0272$. The conditional Triton hot source occupies the same separation between mechanical and thermal eligibility (Fig.~\ref{fig:state}a).

A shallow terrestrial source analog with 2.3 km path, 10 mm grains and 3 km catchment has $\Pi_P=2.64$, $\Pi_H=0.00553$ and $\Pi_V=357$. With 30 km focusing its heat margin rises to 0.0590. Every sampled positive-loaded steady cold dry-source case remains heat-limited in this comparison. A separate full source--geometry--stress screen of 10,080 designs gives 4,834 cases satisfying the relaxed opening condition and 72 cases with an optimistic heat margin above or equal to unity at the unloaded segregation rate, but none above unity at the actual steady hydraulic rate; no case satisfies the joint conditions (SI Table~\ref{tab:joint}). Pressure and unloaded melt flux by themselves thus overstate continuous cold transport.

The matched-pulse experiment instead isolates path length and pressure at identical high input rate and finite inventory. A 5 million m$^3$, 46 h, 1550 K input (a terrestrial event-scale design motivated by Axial eruption volume and duration proxies \citep{lesaout2020}) is stipulated, with mechanically held aperture. For 100 m strike and 0.5 m aperture, the 1.1--4.5 km terrestrial distances give $\Pi_H=12.3$--37.7 and $\Pi_V=10.5$--16.4. The 171 km Europa path reaches $\Pi_H=0.965$ and $\Pi_V=0.569$; the approximately 89 km early-hot path has $\Pi_H=1.57$ and $\Pi_V=1.07$. The nominal 149 km Triton path has sufficient optimistic heat ($\Pi_H=1.05$) but insufficient fill-plus-payload material ($\Pi_V=0.648$). The same large pulse can thus move from strong thermal and inventory margins to near or below the transport boundary as path length, cold boundary and hydrostatic pressure change (Fig.~\ref{fig:state}b). Smaller widths or shorter warm paths shift those boundaries. The supplied pulse is a controlled finite input, distinct from the predicted self-supplied rates.

\subsection{Time-dependent propagation and thermal arrest}
\label{sec:front}
\begin{figure}[!htbp]
\centering\includegraphics[width=\textwidth]{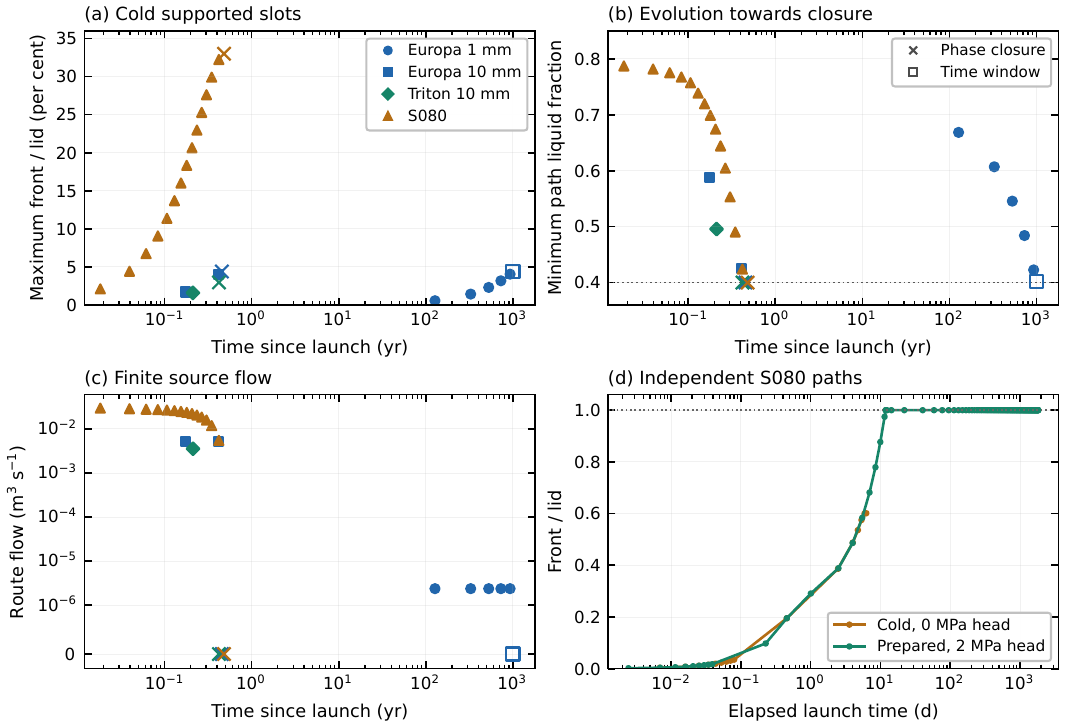}
\caption{Time-dependent propagation through cold rock and conditional escape from thermal arrest. (a)--(c) Actual stored positive-time checkpoints of the same four 1024-cell common-model runs, with 100 m by 0.1 m held slots and no prior preparation. The maximum wet-length fraction, minimum path liquid fraction and route rate share the launch clock. Crosses mark resolved thermal phase closure; open terminal squares mark the 1000 yr time window of the 1 mm Europa run. The dotted line in (b) is the connectivity criterion, $f_l=0.4$. The flow at every terminal checkpoint is recorded as zero; the symmetric-log ordinate in (c) retains these points and the low fine-grain flux. S080 is a separate reconstructed source applied to the common model. (d) Independent S080 advancing-slit solver, $W=1$ m, $b=0.5$ m: cold route with zero receiver head and a separately prepared route with 100 yr prior contact and 2 MPa receiving head. The launch clock excludes the paid preparation and includes only actual saved evolution checkpoints. These paths have distinct thermal and receiving conditions from panels (a)--(c). The horizontal line denotes the full lid.}
\label{fig:front}
\alttext{In the first three panels, cold-slot fronts advance only partway through the rock, with three histories reaching the liquid-connectivity threshold within a year. The fine-grained Europa case advances slowly over 1000 years. In the fourth panel, the independent cold S080 slit reaches about 60 per cent of the lid, whereas the prepared path reaches the bed.}
\end{figure}

The transient calculation tests the heat limitation while advancing actual magma and resolving changing liquid connectivity. For representative cold held slots with 100 m strike and 0.1 m aperture, keeping the mechanically supported geometry fixed isolates the thermal fate of the molten route. At the finest tested axial resolution (1024 cells), maximum wetted length is 4.36 per cent of the lid in the hot Europa 1 mm case, 4.40 per cent at 10 mm, and 2.96 per cent in the hot conditional Triton 10 mm case. The independently reconstructed low-gravity S080 source imported into the common model advances 33.01 per cent of its cold lid. None delivers bed material (Fig.~\ref{fig:front} and SI Table~\ref{tab:u_front_ref}). The actual saved rates and liquid fractions explain their different time histories: the coarse Europa and Triton examples reach the phase-connectivity limit after 0.455 and 0.421 yr, while S080 reaches it after 0.481 yr. The fine Europa example advances slowly, ending at the 1000 yr calculation window with minimum liquid fraction 0.4022, still marginally above the 0.4 threshold. This last case is time-limited rather than a recorded phase-closure event.

Independent finite-batch and sustained-feed calculations complement the common geometry. The three reconstructed detached finite batches traverse fractions 0.326--0.750 of their corresponding cold lids, while insulated controls reach the full length (SI Section~\ref{si:reference_controls}). For the separate S080 held 0.5 m by 1 m slot, the cold continuous-feed route reaches 0.601 of the lid before liquid arrest. With prior 100 yr of specified thermal contact, the same source and supported dimensions allow complete transit and a 5 million m$^3$ bed payload against a 2 MPa receiver, after about 4.99 yr of launch and discharge evolution (Fig.~\ref{fig:front}d). The preparation alone costs $1.14\times10^{16}$ J or 8.45 million m$^3$ of sacrificed material; the subsequent sustained rate is about 0.03193 m$^3$ s$^{-1}$. These are distinct time and energy expenditures. Together with the cold and insulated finite-batch results, they identify thermal connectivity as a dynamically consequential restriction even with mechanical access. Additional endpoint and grid tests, including focused catchments, are reported in SI Sections~\ref{si:reference_controls} and \ref{si:u_numerics}.

\section{Recurrent delivery and thermal resetting}
\label{sec:u_recurrence}
\begin{table}[!htbp]
\caption{Thermal resetting and repeated-delivery budgets over 0.1 Myr using the same Gaussian receiver. First-rest pressure $p_c/p_\infty$ and upper-wall temperature $T_{\rm wall}$ are from the 5 kyr paid-access branch immediately before renewed contact; its minimum liquid fraction is zero. The next four columns give actual cumulative bed arrival ($10^6$ m$^3$) under the specified 5 and 10 kyr opportunity schedules, with target 5. $V_p$ is total sacrificed magma-equivalent heating volume for the 5 kyr paid case. Radii in the source labels are catchment radii. The three branches share the same specified initial preparation.}
\label{tab:reset}
\centering\small\setlength{\tabcolsep}{3pt}
\begin{tabularx}{\linewidth}{@{}>{\raggedright\arraybackslash}Xrrrrrrr@{}}
\toprule
 & & &\multicolumn{4}{c}{Bed arrival ($10^6$ m$^3$)}&\\
\cmidrule(lr){4-7}
Source & $p_c/p_\infty$ & \shortstack[r]{$T_{\rm wall}$\\(K)} & \shortstack[r]{Sealed\\5 kyr} & \shortstack[r]{Adjacent\\5 kyr} & \shortstack[r]{Paid\\5 kyr} & \shortstack[r]{Paid\\10 kyr} & \shortstack[r]{$V_p$\\(km$^3$)}\\
\midrule
\AResetRows
\bottomrule
\end{tabularx}
\end{table}

\begin{figure}[!htbp]
\centering\includegraphics[width=\textwidth]{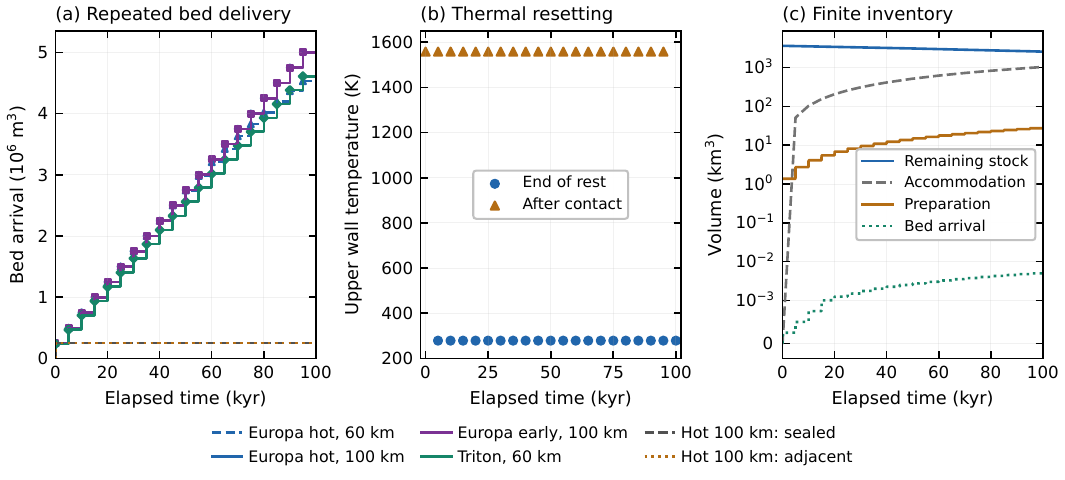}
\caption{Recurrent source--path evolution for the 5 kyr schedule with a common Gaussian receiver. Twenty specified opportunities of at most 46 h delivery follow specified initial preparation, and the paid branch has 40 yr of renewed contact before later opportunities. The path uses a held 100 m by 0.2 m slot, a 3 km collector and the indicated 60 or 100 km catchment. (a) Cumulative actual bed arrivals at recorded event endpoints for four paid histories; sealed and cold-adjacent hot Europa 100 km controls both deliver only on the first opportunity and lie on the same low curve. Steps indicate cumulative event totals. (b) Actual upper-wall temperatures of the hot Europa 100 km paid history at the end of each rest and after each contact. (c) Remaining accessible stock, cumulative accommodation, sacrificed magma-equivalent preparation and actual bed arrival on the same hot Europa 100 km paid history, on a symmetric-log volume axis. All times include preparation and rest, including final dormancy to 0.1 Myr. The mechanically accessible initial route and subsequent contacts are stipulated; their cost is incurred in the budgets. A distinct mapped-receiver history is reported in SI Section~\ref{si:recurrence}.}
\label{fig:repeat}
\alttext{Repeated paid contact progressively delivers about 4.5--5 million cubic metres, while the sealed and cold-adjacent controls remain near 0.25 million cubic metres. The upper wall repeatedly cools to near-ocean temperature and is reheated. The finite-source budget shows that accommodation and preparation consume much more magma than reaches the bed.}
\end{figure}

\subsection{Recharge and thermal-reset timescales}
\label{sec:clocks}
The long path requires thermal connectivity not just during initial arrival but also between episodes of transport. Let
\begin{equation}
t_{\rm cool}=\frac{W^2}{\pi^2\kappa_r},\qquad
t_{90}=C_e\int_0^{0.9p_\infty}
\frac{\dd p}{Q_s(p)-Q_a(p)} ,
\label{eq:clocks}
\end{equation}
where $\kappa_r=k_r/(\rho_rc_{p,r})$ is the host thermal diffusivity. The former is the first finite-strike diffusion timescale and the latter the time for a sealed collector with frozen-date source properties to charge from zero to $0.9p_\infty$. The repeated-access experiments use $W=100$ m for both thermal and route geometry, $b=0.2$ m and a 3 km collector. Hot Europa and conditional Triton examples have $t_{\rm cool}=40.9$--41.4 yr and $t_{90}=2.88$--9.63 yr. Their first inter-event rest of about 4,960 yr before renewed contact gives $\Pi_T\simeq120$ and $t_{\rm dorm}/t_{90}=515$--1723 (Fig.~\ref{fig:state}c). The same illustrative diffusion width at the terrestrial 4 and 13 yr Axial eruption intervals gives $\Pi_T=0.10$ and 0.33, indicating a different separation of timescales at those intervals \citep{yang2024}. This comparison uses the same specified width; the diffusion time scales quadratically with $W$.

The evolved source and wall states independently confirm the time-scale interpretation. During the first rest of the coupled 5 kyr histories, collector pressure returns to almost its initial relaxed value (0.9982--1.0009 times $p_\infty$), while the upper wall returns to 275--282 K and the minimum path liquid fraction is zero (Table~\ref{tab:reset}). The collector and path therefore retain different memories of the previous episode. Pressure recovery is compatible with complete loss of upper-path liquid continuity.

\subsection{Conditional renewed access and finite resource budgets}
\label{sec:repeat}
We compare three inherited-path states after a common initial preparation: sealed subsequent access, a cold adjacent trace, and renewed thermal contact for 40 yr with the mechanically accessible held route. During contact the wall is warmed toward the diffusion profile consistent with the local rock temperature and inlet enthalpy. The integral of incremental wall heating plus side losses gives $E_p$; the sacrificed preparation volume and fill volume are debited from the contemporaneous source. The contact calculation tests the same finite stock, charging, net supply and heating time for both costs. Its definitions and inherited plug/refill ledger are given in SI Section~\ref{si:recurrence}. The specified contact permits a new thermally eligible path only where the available finite resource satisfies both requirements (Fig.~\ref{fig:state}d).

The initially prepared Europa routes with sealed subsequent access deliver 0.25 million m$^3$ each in the first opportunity (0.235 million m$^3$ for conditional Triton). Their subsequent delivery ceases even as the source recharges; the cold-adjacent controls give similar totals. In contrast, repeated paid thermal contact allows continuing transfer from the finite source (Fig.~\ref{fig:repeat}). With twenty opportunities separated by 5 kyr, the two 100 km focused Europa thermal cases attain the 5 million m$^3$ (0.005 km$^3$) target. Preparation requires 27.39 and 14.39 km$^3$ of magma-equivalent material, respectively, about 5,480 and 2,880 times their bed-arrival volumes. The 60 km Europa and conditional Triton source cases deliver 4.529 and 4.608 million m$^3$, respectively. Keeping the 0.1 Myr horizon while changing the specified interval to 10 kyr alters cumulative delivery and preparation costs (Table~\ref{tab:reset}). All these comparisons use the same Gaussian receiving model.

The inventory history of the hot Europa 100 km Gaussian paid case shows the partition that underlies the small delivered fraction. Of the initial 3655.5 km$^3$ accessible inventory, approximately 1040.6 km$^3$ is accommodated within the rock during the calculation; 2587.3 km$^3$ remains, 27.39 km$^3$ is sacrificed for preparation, and 0.005 km$^3$ enters the ocean. The remaining volume terms in equation (\ref{eq:ledger}) include elastic storage, the active path and old plugs. The upper-wall temperature repeatedly resets from the hot prepared state to the seafloor thermal boundary (Fig.~\ref{fig:repeat}b,c). Preparation and the magma left in earlier plugs have separate material and enthalpy ledgers (SI Section~\ref{si:recurrence}). Thus a pressure-generating and still substantial finite source can support renewed delivery when favourable geometry and time are available, yet pressure recovery alone does not provide the material and energy needed to restore the path.

\FloatBarrier
\section{Discussion}
\label{sec:discussion}
\subsection{Icy-world transport regimes and relation to previous work}
The coupled calculations identify three different selectors of silicate volcanism. A planet or satellite can have a molten region separated from its water layer by solid high-pressure ice; a directly liquid-contact interior can lack connected melt or positive dry-magma buoyancy; and a dense, molten source can develop sufficient fracture pressure while remaining unable to sustain a liquid path to the ocean. The third selector is particularly consequential for deep melting on icy worlds. Although buoyant segregation, creep and wall freezing also occur on Earth, weak gravity and grain-scale rheology determine whether pressure can first accumulate, while long cold travel distances and an ice-loaded cold ocean impose independent thermal and inventory demands. The separation of these restrictions emerges from the combination of planetary conditions, rather than from a single gravity scaling.

The results extend three complementary strands of Europa research. The mantle thermal histories of \citet{behounkova2021} establish the feasibility of melting for favourable heating epochs; our fixed dated source calculations connect such thermal states to the finite material and enthalpy reaching an ocean. \citet{bland2022} quantify the challenging injection and fracture propagation requirements for Europa, including the high-flux endmember of order 1 m$^2$ s$^{-1}$ per unit strike and the possibility of episodic accumulation. \citet{green2025} identify small melt fractions and low intergranular magma supply that strongly restrict driving in their present-day Europa structures; their computed dike influxes are about four orders of magnitude below the eruption requirement. The independently evaluated published trajectories provide a comparison of penetration (SI Section~\ref{si:external}). The grain-scale and stress-dependent pressure reversal in our matched sources exposes an additional regime: even when relaxed pressure suffices for an inherited fracture, the same finite source can lack the heat, time or inventory needed to turn that opening into continuous ocean access. Its geological repetition adds a memory constraint independent of the initial thermal and mechanical state.

The terrestrial-distance and fixed-packet controls clarify the origin of this difference. A dry self-supplied cold Earth analog can itself fail a large heat requirement. The established, frequently reactivated warm routes of active terrestrial volcanic systems occupy different thermal states. Holding the cold starting geometry constant instead isolates the additional heat demand arising from path lengths of tens to hundreds of kilometres on the selected icy worlds. With the same supplied pulse, the early-hot Europa and longer contemporary-like Europa profiles straddle the joint optimistic heat and volume thresholds. High absolute pressure further shifts near-solidus phase state at fixed geometry and input; an icy world's ocean boundary therefore matters to both delivery and subsequent emplacement.

\subsection{Implications for ocean--rock exchange and planetary evolution}
Interior heating and ocean access regulate different components of planetary evolution. The finite-source ledger shows that even a relatively large accessible reservoir can retain most of its material in the interior, or expend enthalpy and extractable magma on heating, filling and restoring pathways. The large preparation-to-emission ratios in the favourable Europa repeated-access experiments provide a quantitative example of this partition. Internal intrusive activity and thermal exchange need not be proportional to the volume of fresh erupted silicate at an ocean floor. During millennial dormancy, a collector can recover its pressure long before the next delivery opportunity, while the transport pathway loses its thermal connectivity. Later events then face a fresh enthalpy and material demand, rather than simply releasing the pressure accumulated since the last event.

Consequently, the residence time of magmatic heat in the rocky interior and the material flux to the ocean depend on the history of accessible transport as well as on melt production. Long-term interior or ocean--rock exchange models may treat source generation and successful bed delivery as distinct coupled fluxes, with material retained at depth and heat expended before arrival. Aqueous circulation remains a separate mechanism of water--rock interaction \citep{choblet2017,sekine2015,fisher2024}. Restricting silicate magma transport is compatible with significant hydrothermal circulation. Equating the chemistry or heat transfer of that circulation with erupted melt volume would conflate processes whose pathways and timescales can differ.

\subsection{Geological accessibility, assumptions and uncertainty}
The calculated domains depend on the rock composition, melt segregation and strain laws, source geometry, inherited temperature profile and receiving boundary. The dry-peridotite phase closure and reference basalt phase interval, grain-size dependence and creep parameters are constitutive choices; hydrated or compositionally distinct lithologies can shift or alter these domains. Likewise, the adopted Enceladus porous-density endmember, conditional liquid-ocean Triton and selected high-pressure ice columns represent different eligible structural classes. The inherited thermal endmembers and their dated evolution constrain the calculations, rather than providing a unique planet-age thermal inversion.

The favourable recurrent histories require focusing over radii of 60--100 km into a 3 km collector, an initially prepared path, mechanical support and the specified 40 yr full-path thermal contact. The budget demonstrates the cost and conditional transport behaviour of these accessible routes. Establishing such a long open contact from a cold, intact lithosphere is an additional geological requirement. The direct mapped cases provide explicit support-demand estimates, with compressive contact fractions up to about 0.153 in the adopted 50 MPa strength model (SI Sections~\ref{si:u_path} and \ref{si:u_numerics}). The large heat requirements quantify why retaining or re-establishing access matters, but the geologic availability of this focusing and support remains uncertain.

The 5 and 10 kyr opportunities and 46 h high-rate comparisons are prescribed experimental clocks; $t_{\rm cool}$ is a finite-strike diffusion scale rather than an exact freeze-closure time. Changes in width and permeability alter both the thermal memory and hydraulic supply; changes in receiver head and evolving composition further modify the joint conditions. The four margins organise these necessary conditions in terms of the matched geometry and target. They are not sufficient conditions for dynamic liquid transport. The independent source and fracture reconstructions, numerical tests and refined supported endpoints strengthen the mechanism-level interpretation, while leaving constitutive and geological accessibility uncertainties distinct from solution discretisation. In particular, attainment of the 0.005 km$^3$ target under a specified recurrent schedule establishes a conditional delivered volume, not a natural maximum volcanic volume.

\section{Conclusions}
\label{sec:conclusions}
Weak buoyancy, deep melt production, an extensive cold rocky lithosphere and the ice-loaded ocean boundary jointly select a distinct magma transport regime on icy ocean worlds. Grain-size-dependent segregation and nonlinear relaxation can reverse the opening-pressure restriction, yet a long fracture may remain thermally arrested or exceed the finite source inventory even after that reversal. High absolute ocean pressure further penalises near-solidus transport in the reference basalt phase law. During geological dormancy, pressure can recover while the liquid path freezes; each renewed delivery then consumes substantial material and enthalpy to restore access. This coupled pressure--heat--inventory--memory regime separates interior melting from seafloor volcanism and makes renewed access to the ocean an independent regulator of recurrent silicate activity on icy worlds.

\section*{Data availability}
The data and code underlying this article, including planetary profiles, source and transport results, histories, finite budgets, independent checks and figure and table generation scripts, are archived on Zenodo at \href{https://doi.org/10.5281/zenodo.24493739}{doi:10.5281/zenodo.24493739} \citep{hestaphin2026data}. The published Green et al. model data and inputs are available at \href{https://doi.org/10.5281/zenodo.10850608}{doi:10.5281/zenodo.10850608} \citep{green2024data}. Planetary parameters follow \citet{jpl2026}; the mapped Axial fissure outlines are available at \href{https://doi.org/10.1594/IEDA/324417}{doi:10.1594/IEDA/324417} \citep{axialoutlines}. Other observational sources are identified by their references.

\FloatBarrier
\bibliographystyle{elsarticle-harv}
\bibliography{references,new_refs}

\clearpage
\newgeometry{margin=24mm}
\makeatletter
\def\@blstr{1}
\makeatother
\renewcommand{\baselinestretch}{1}
\fontsize{11}{13.6}\selectfont
\setlength{\tabcolsep}{5pt}
\renewcommand{\arraystretch}{1.12}
\renewcommand{\topfraction}{.92}
\renewcommand{\bottomfraction}{.88}
\renewcommand{\textfraction}{.1}
\setlength{\parskip}{.25em}
\renewcommand{\thepage}{S\arabic{page}}
\setcounter{page}{1}
\begin{center}
{\large\bfseries Supplementary Information\par}
\vspace{.5em}
{\large Weak buoyancy and cold lithospheres decouple interior melting from seafloor volcanism on icy ocean worlds\par}
\vspace{.5em}
Nahiz Hestaphin\\[2pt]
{\small Correspondence: \href{mailto:zh389@cantab.ac.uk}{zh389@cantab.ac.uk}}
\end{center}
\setcounter{section}{0}\setcounter{subsection}{0}
\renewcommand{\thesection}{\Alph{section}}
\renewcommand{\theHsection}{supp.\Alph{section}}
\setcounter{figure}{0}\renewcommand{\thefigure}{S\arabic{figure}}\renewcommand{\theHfigure}{supp.\arabic{figure}}
\setcounter{table}{0}\renewcommand{\thetable}{S\arabic{table}}\renewcommand{\theHtable}{supp.\arabic{table}}
\setcounter{equation}{0}\renewcommand{\theequation}{S\arabic{equation}}\renewcommand{\theHequation}{supp.\arabic{equation}}
The derivations and controls below specify the shared transport model, its heat and material budgets, independent source/fracture tests and the numerical range of its conclusions. Unprefixed figures, tables, equations and numbered sections refer to the main paper; figures, tables and equations prefixed S and lettered sections refer to this supplement. The evidence index and case-level data are provided in the accompanying Zenodo archive \citep{hestaphin2026data}.

\section{Mass-constrained planetary columns and correlated thermal states}
\label{si:u_structure}
The common radial fields obey equation (\ref{eq:u_hydro}) of the main paper. The selected structures and their boundary pressures are given in main Table~\ref{tab:u_worlds}.
The radius and gravitational parameter are taken from the JPL observations \citep{jpl2026}. The layer and interface conditions use the references \citet{vanceinterior2018,park2024,gaeman2012}; liquid thermodynamics and phase boundaries follow \citet{iapws2018,iapwsmelt}. The historical and slush Titan cases refer to \citet{petricca2025}.
The planetary calculation integrates hydrostatic pressure and enclosed mass from the outer radius down to the rock interface. Ice Ih has reference density 920 kg m$^{-3}$. Liquid density is evaluated with the same validated IAPWS-95 cold-liquid equation used in the reference-water comparisons. For ice III, V and VI we use rounded compressible solid endmembers \(\rho=\rho_0(1+4P/K_0)^{1/4}\), with \(\rho_0=1140,1200,1270\) kg m$^{-3}$ and \(K_0=12,13,14.05\) GPa. These are reduced representations of the phase-dependent compression documented by \citet{bezacier2014,journaux2020}, with phase-specific density rather than a full multicomponent Gibbs-energy closure. The layered geometry follows the corresponding published reference scenario \citep{vanceinterior2018}; temperature across the liquid layer is interpolated from a point just above the melting temperature at its top to the specified lower temperature. The minimum liquid melting margin is tested through the depth profile.

Mass and moment are evaluated independently. For layers of constant density between \(r_1,r_2\), their mass and polar moment are \(4\pi\rho (r_2^3-r_1^3)/3\) and \(8\pi\rho(r_2^5-r_1^5)/15\). The Europa two-layer rock fits the mass and hydrostatic moment at the reference metallic-core radius. Ganymede retains the published FeS-bearing inner endmember of density 7203 kg m$^{-3}$ and fits a common high-pressure ice density factor to its moment; the fitted correction and layer density/pressure fields are archived. Callisto and historical Titan mixed-interior scenarios use a mass-and-moment fit with an assumed central region where the resulting bulk densities are admissible. The homogeneous alternative reports its actual moment. The historical shallow Europa 15+60 km column supplies the useful independent gravity-matching control, but does not satisfy the same moment at plausible densities. The cold historical Titan ocean geometry has a moment discrepancy within the present reduced density closure; the other historical moment and the illustrative slush branch are separately reported. A direct-contact calculation uses only the indicated liquid geometry and moment compatibility (or Triton's conditional geometry where a moment is not independently constrained). Solid basal layers remain a separate interface class.

\begin{table}[H]\centering\footnotesize
\caption{Planetary structure audit. Densities are effective bulk layer densities in kg m$^{-3}$; $C_I/MR^2$ is the mass-normalized moment. The calculated pure-water minimum melting margin tests the liquid path, not the composition of an actual ocean. All structures are mass-closed.}
\label{tab:u_structaudit}
\begin{tabularx}{\textwidth}{@{}lrrrrrrX@{}}\toprule
Column&$\bar\rho_r$&$\rho_r$&$\rho_{\rm inner}$&Moment&Target&Margin (K)&Interface\\\midrule
\UStructureRows
\bottomrule\end{tabularx}
\end{table}

Thermal evolution of the rocky layer uses the same absolute pressure and gravity from the spherical structure. The thermal cell volume is \(4\pi(r_{i-1/2}^3-r_{i+1/2}^3)/3\), the corresponding surface area is \(4\pi r^2\), and the two face fluxes and allocated volumetric rock heating determine the change in its total sensible/latent energy. The surface conductive flux and the integrated heat-source term close the rock energy ledger. The dry melting reference uses the Katz et al.\ dry solidus \(T_s=273.15+1085.7+132.9P_G-5.1P_G^2\) (temperatures expressed numerically in K and \(P_G=P/(1\,\mathrm{GPa})\)), lherzolite liquidus scale \(T_l=273.15+1475+80P_G-3.2P_G^2\), exponent 1.5 and the clinopyroxene exhaustion fraction \(0.17/(0.5+0.08P_G)\). The remaining higher-fraction branch is joined continuously to the dry liquidus endmember \(273.15+1780+45P_G-2P_G^2\). The connected segregation interval is restricted to \(P_G\leq 8\) and \(\phi_m=0.005\)--0.225. Composition follows the specified dry endmember.

The spherical thermal enthalpy equation (\ref{eq:u_rock}) uses a fixed bed temperature and an insulating inner boundary. The melting law follows \citet{katz2003}.

\begin{table}[H]\centering\small
\caption{Specified initial rock thermal histories. Rock radiogenic powers for Europa and Enceladus are literature-motivated reference allocations; Triton's 50 GW is a declared radiogenic scale. Additional tidal power is assigned only in the specified early/hot limits. These histories are finite-time checks of inherited thermal states, not a whole-age convection inversion.}
\label{tab:u_thermalset}
\begin{tabularx}{\textwidth}{@{}Xlrrr@{}}\toprule
Thermal state&Column&$T_{\rm int}$ (K)&Initial lid (km)&Rock power (GW)\\\midrule
Europa cold&Eu\_V18\_cold&1450&250&100\\
Europa hot&Eu\_V18\_cold&1560&180&100\\
Europa early hot&Eu\_V18\_warm&1600&100&200\\
Enceladus cool&En\_P24&650&60&0.27\\
Enceladus dry sensitivity&En\_P24&1450&45&0.27\\
Enceladus alternative dry&En\_P24\_alt&1480&60&2.27\\
Triton cool&Tr\_liquid&1250&250&50\\
Triton hot&Tr\_liquid&1550&160&50\\
\bottomrule\end{tabularx}
\end{table}

The spherical enthalpy integration uses 1,536 thermal cells and a 25 kyr maximum interval over reference ages 0, 0.1, 1 and 10 Myr. The initial temperature increases linearly from the rock-interface temperature through the stipulated lid to the interior value. The independent finer histories and their source reconstructions are in the numerical archive. The long-term coupled test fixes its source interval to the overlap of the connected depth intervals at 0 and 0.1 Myr. The constitutive profiles evolve between those dated fields; the stock is the smaller accessible liquid volume and is not reset. Local melt extraction is modeled as removal of at most 2 percent of the inherited connected-melt material, with the evolving host temperature as its source enthalpy. Local thermal exchange required to maintain the collector enthalpy is recorded. The extraction energy above bed temperature is also compared with the finite accessible enthalpy and local rock thermal inventory. For the cool hydrated-core context of Enceladus see \citet{choblet2017,sekine2015}.

\FloatBarrier
\section{Nonlinear segregation and collector rheology}
\label{si:u_source}
Equations (\ref{eq:darcy}), (\ref{eq:u_comp}) and (\ref{eq:u_collect}) of the main paper define the nonlinear melt flux, matrix closure and elastic collector. Main Table~\ref{tab:u_source} reports the finite source states. The common time-dependent histories evolve the fixed physical source interval between the dated thermal profiles; depletion reduces the source curve.

The compaction law uses an effective scalar stress invariant and parallel dry-olivine mechanisms. With absolute pressure $P$, nonnegative scalar stress invariant $\sigma$, grain scale $d$, melt fraction $\phi_m$ and molar gas constant $R_g$,
\begin{align}
a_d&=A_dd^{-3}\exp\left[-\frac{E_d+PV_d^{\rm act}}{R_gT}+20\phi_m\right],\\
a_s&=A_s\exp\left[-\frac{E_s+PV_s^{\rm act}}{R_gT}+20\phi_m\right],\\
\dot\epsilon_{\rm II}&=a_d\sigma+a_s\sigma^{3.5}.
\end{align}
The reference coefficients are $A_d=1.5\times10^{-15}$ m$^3$ Pa$^{-1}$ s$^{-1}$, $A_s=1.1\times10^{-16}$ Pa$^{-3.5}$ s$^{-1}$, $E_d=375$ and $E_s=530$ kJ mol$^{-1}$, $V_d^{\rm act}=8.2\times10^{-6}$ and $V_s^{\rm act}=17\times10^{-6}$ m$^3$ mol$^{-1}$. These diffusion/dislocation constants are those summarized by \citet{schulz2020} from \citet{hirth2004}. Melt weakening is a declared exponential closure; sensitivity to grain size, material stress and spatial rheology is retained. The Darcy drag uses \(k_\phi=d^2\phi_m^3/100\), with a minimum modeled porosity in the drag/compaction operator of 0.002. The source interval begins at the continuous \(\phi_m=0.005\) crossing. Cosine-spaced control volumes resolve the low-porosity boundary layer. For each trial collector traction, the cell-centered \(q=(B+c')/(\mu_l/k_\phi)\) is eliminated in favor of face traction \(c\); the interfacial relation \(q_{i+1}-q_i=\Delta z\,2(a_d\phi_m c+a_s|\phi_m c|^{2.5}\phi_m c)\) yields a nonlinear tridiagonal Newton system. The boundary conditions are prescribed top traction and zero basal flux. Monotone pressure interpolation uses 33 resolved traction values; the pressure where top flux vanishes is localized by a scalar root.

For the cavity-pressure integral (equation \ref{eq:cavity}), the radial geometry sets the local strain rate. The same strain rate is imposed at each angular position and the nonlinear stress is inverted before averaging. Hot and cold directions are additional endmembers. The outer radius is sampled at 4, 16 and 64 cavity radii, truncated so the spherical sampling remains inside the rock; this remains a reduced radial-deformation field. In the homogeneous Newtonian limit \(p=\eta Q_a[1-(R_c/R_{\rm out})^3]/(\pi R_c^3)\). Elastic storage uses \(K_l=10\) GPa, \(G_r=30\) GPa. The creep flux uses monotonic logarithmic interpolation of the computed cavity curve.

The source-reversal condition at any pressure \(p_{\rm req}\) is both \(p_{\rm req}<P_*\) and \(Q_s(p_{\rm req})>Q_a(p_{\rm req})\) before transport is allowed; subsequent transport adds \(Q_f\) to the outflow. For the 1 mm hot Europa source, at 0.05 MPa the creep rate is about ten times the loaded inflow; at 0.2 MPa it is about 73 times the inflow. The corresponding reduction of accommodation would be required to support charging to those pressures. A 2 MPa requirement exceeds that source's stopping pressure even without accommodation. In the complete tested spatial/creep matrix with 0.3--10 mm grains, 0.1--10 MPa imposed compaction stresses, three cavity scales and angular/outer-radius endmembers, the Europa hot relaxed pressure spans approximately 3.7 Pa to 19.35 MPa. These are constitutive endmembers, not probabilities of specific mantle states. The matched dry-gate and continued-feed cases use the fully nonlinear compaction curve at its self-generated stress.

\section{Fracture initiation, connectivity and cold-wall demand}
\label{si:u_fracture}
\subsection{Mechanical gates}
\begin{figure}[!htbp]\centering\includegraphics[width=.91\textwidth,height=.59\textheight,keepaspectratio]{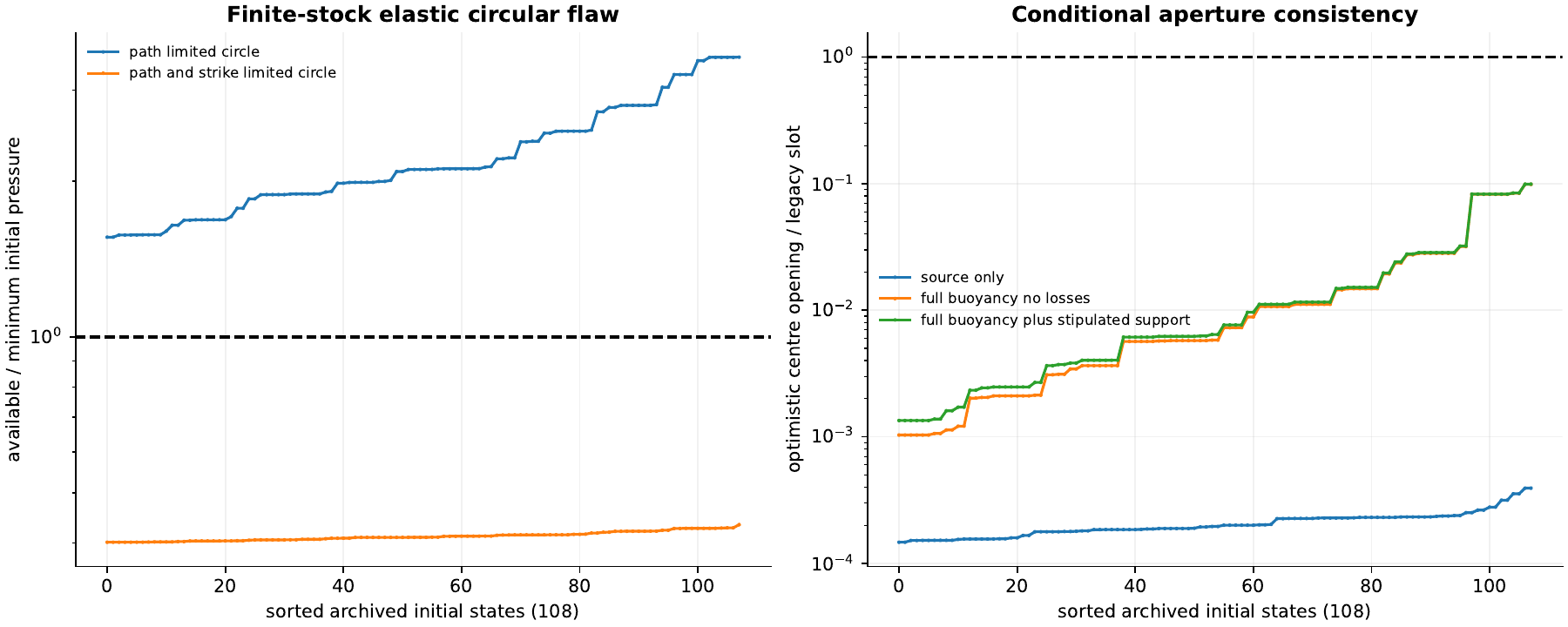}
\caption{Finite-source elastic opening against circular flaw and elongated slot requirements. Circles restricted by the strike width have a larger opening demand than favorable path-limited circles. The aperture ratio refers to a specified slot; unity marks the necessary opening condition.}\label{fig:elastic}
\end{figure}

The elongated-slot opening screen is equation (\ref{eq:u_open}) of the main paper. A held connected slit still requires adequate mechanical support. The fissure segments clipped to the adopted bathymetrically supported Axial H3 outline total approximately 296 m exposed strike \citep{axialoutlines,lesaout2020}; the wider geometries are controlled deep-route endmembers.
Equation~\ref{eq:u_open} separates elastic aperture, seed toughness and tensile strength. For a fluid-accessible circular flaw there is an independent pressure--volume comparison: with radius $a$, $p_K=K_I\sqrt\pi/(2\sqrt a)$ and elastic compliance $C_f=16a^3/(3E')$, the necessary collector pressure is
\[
p_{\rm req}=[p_K-\sigma_T-2Ba/3]_++C_fp_K/C_e,
\qquad B=(\rho_r-\rho_l)g.
\]
The last term finances the liquid volume of the flaw. The upper center aperture satisfies $b_{\max}=8a[p_c+\sigma_T+Ba]_+/(\pi E')$ under favorable complete liquid access. It is a different geometry from the elongated held slit in Eq.~\ref{eq:u_open}. Independent elastic-opening comparisons using the reconstructed source states are shown in Fig.~\ref{fig:elastic}. Bounds use 1/2.4 MPa$\sqrt{\rm m}$ toughness, 3/8.5 MPa intact strength and widths constrained by the strike length \citep{davis2020,balme2004,goehring2008}.

\subsection{Crystallization and halfspace heat integral}
Equations (\ref{eq:u_h}) and (\ref{eq:basalt_visc}) specify the bounded pressure-dependent basalt phase and viscosity law ($\mu_0=20$ Pa s), with loss of liquid connectivity at $f_l<0.4$. The constitutive comparison varies solidus 1250--1375 K, interval width 120--230 K, pressure slope 0--100 K/GPa, latent heat $(3$--$5)\times10^5$ J kg$^{-1}$ and reference viscosity 10--100 Pa s. Across the 1,620 archived constitutive designs at the specified source/slot geometries, 324 satisfy the pressure gate but the maximum steady first-fill heat ratio is only 0.00386 and none passes the joint gate. Source material retains its independently constructed source curve in this comparison \citep{giordano2008,costa2009,krieger1959}. The mechanical design varies strike 10--10,000 m, aperture 1 mm--0.5 m, seed half-length 10--1000 m and tensile stress 0, orbital scale, 0.5 or 2 MPa. The orbital estimate is $3\rho n^2eR_r^2$; $E'=70/(1-0.25^2)$ GPa, $K_{Ic}=1$ MPa$\sqrt{\rm m}$ and nominal dry strength 2 MPa.

\begin{figure}[!htbp]\centering\includegraphics[width=\textwidth]{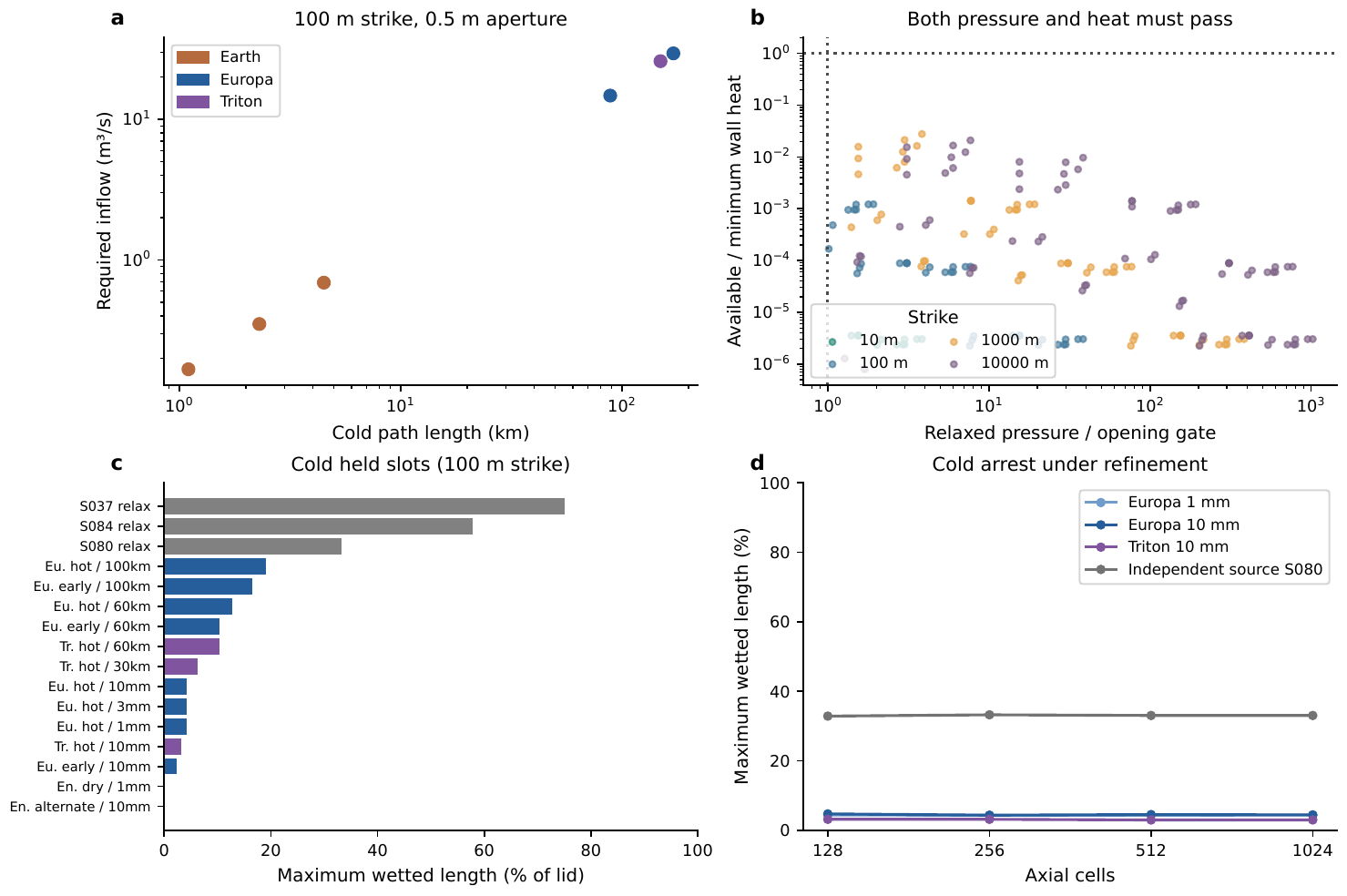}
\caption{Additional panels and refinement controls for main Figs.~\ref{fig:drivers}--\ref{fig:front}. (a) Matched first-fill heat requirement at terrestrial and icy-world path lengths, as in main Fig.~\ref{fig:drivers}b. (b) Opening-pressure and steady heat ratios for inherited traces with fixed 1 km seed and zero additional tension (positive available rate cases). (c) Maximum advance in cold held slots, 0.1 m aperture or 0.5 m in focused catchments. (d) Axial/wall refinement; the bed is at 100 percent. S080 is an independently reconstructed low-gravity source.}
\label{fig:u_thermal}\end{figure}

A wall patch at background $T_{\rm wall}$ in contact with liquid at minimum connected temperature has $\Delta T=[T_{\rm connect}-T_{\rm wall}]_+$, contact conductance $h_c=2k_r/b$ and cumulative heat uptake per area
\[
E_A(t)=\frac{k_r^2\Delta T}{h_c\kappa_r}
\left[\operatorname{erfcx}(\beta)-1+\frac{2\beta}{\sqrt\pi}\right],
\quad \beta=\frac{h_c\sqrt{\kappa_rt}}{k_r}.
\]
Here $E_A$ has units J m$^{-2}$, $\kappa_r=k_r/(\rho_rc_{p,r})$ and $\operatorname{erfcx}(u)=e^{u^2}\operatorname{erfc}(u)$. For first fill, integrate both wetted walls with exposure $t=Wbz/Q_f$ (distance $z$ back from the outlet) and compare with the favorable available heat $\rho_l WbL[h_{\rm in}-\min_z h(T_{\rm connect}(z),P(z))]_+$. This gives the cold-fill rate in Fig.~\ref{fig:u_thermal}a and the heat ratio in panel b. The initial heat screen uses the unloaded rate $Q_0$; for joint transport, wall heat is recalculated at the source- and pressure-limited isothermal hydraulic rate. Every joint design also checks pressure, slot support, finite stock and time. Table~\ref{tab:joint} records the separate classifications.

\begin{table}[H]\centering\small
\caption{Joint steady cold first-fill screening, over all source, aperture, width, seed and stress combinations. Counts describe physical design points. The full design includes source stock, support and hydraulic-time constraints.}
\label{tab:joint}\begin{tabularx}{.80\textwidth}{@{}Xr@{}}\toprule
Requirement & Number of design points\\\midrule
All mechanical and source design points&10,080\\
Relaxed pressure sufficient for the opening screen&4,834\\
Optimistic heat ratio at unloaded source rate $\geq1$&72\\
Heat ratio at actual steady hydraulic rate $\geq1$&0\\
Joint pressure, heat, stock, time and flow conditions&0\\\bottomrule
\end{tabularx}\end{table}

\begin{figure}[!htbp]\centering\includegraphics[width=.97\textwidth,height=.57\textheight,keepaspectratio]{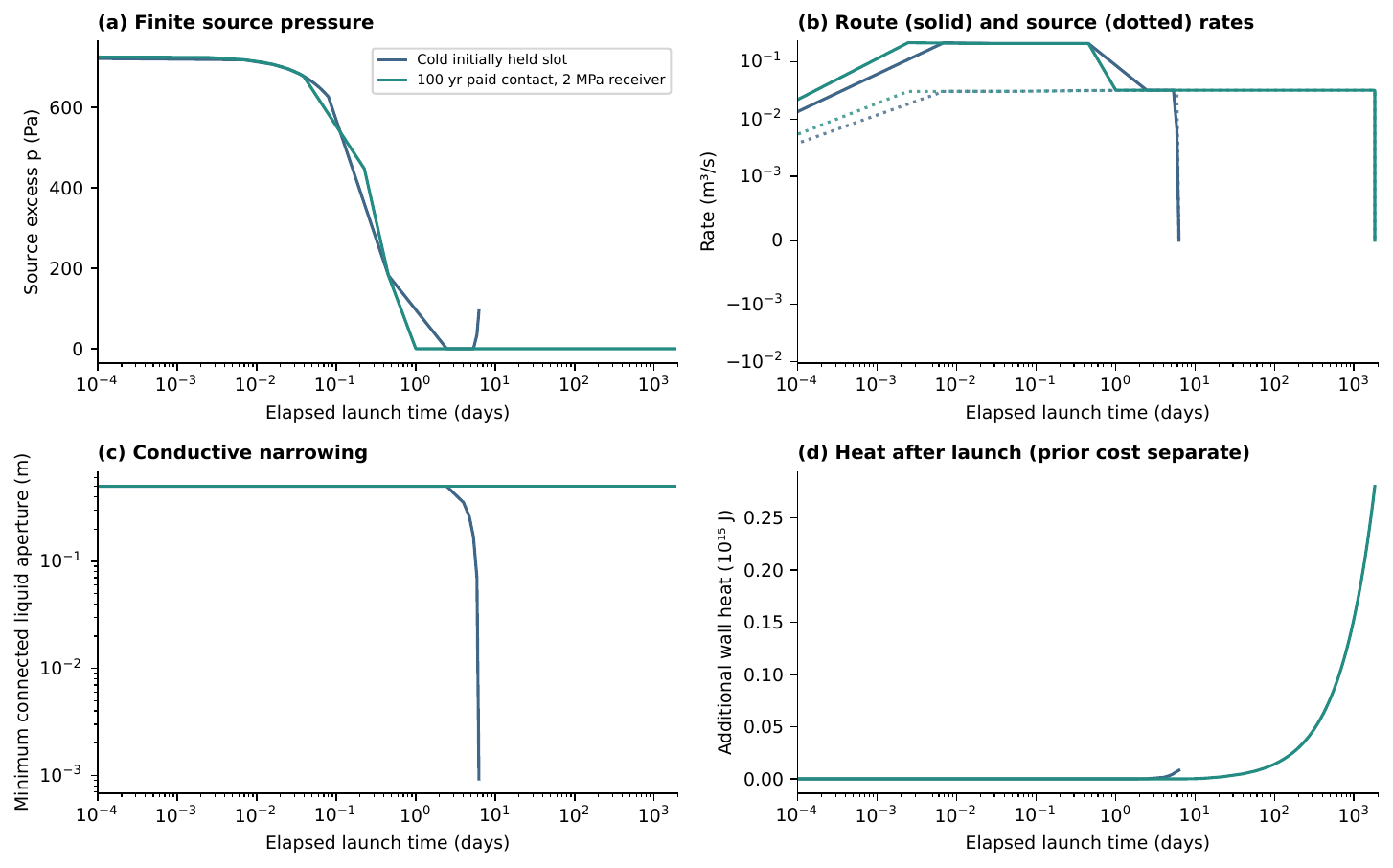}
\caption{Independent time-dependent source pressure, rate, liquid aperture and heat for the same relaxing S080 source (SI~\ref{si:reference_controls}) and held 0.50 m by 1 m slit. The cold route terminates within the lid; the separately paid 100 yr preheated route delivers to a 2 MPa receiver. Prior heat expenditure is distinct from post-launch losses.}\label{fig:rate_aperture}
\end{figure}
\begin{figure}[!htbp]\centering\includegraphics[width=.94\textwidth,height=.57\textheight,keepaspectratio]{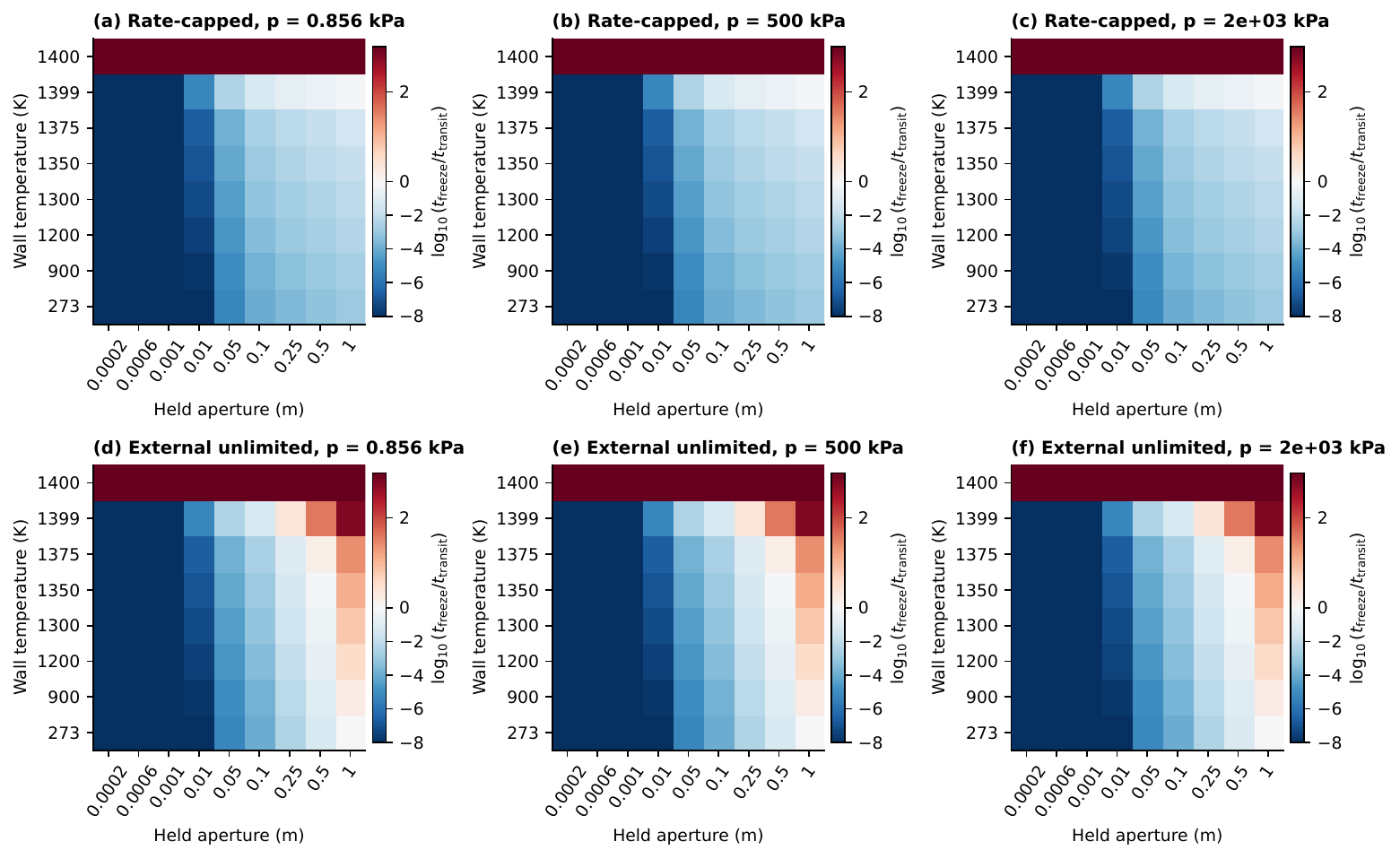}
\caption{Wall temperature, pressure and aperture survival regions for the independently reconstructed S084 source at 10 m strike. The top row uses the actual finite source-rate cap; the bottom row an external unlimited rate under the indicated pressure and connected buoyant column. Positive log closure/transit ratio meets the favorable first-pass thermal screen.}\label{fig:thermal_slices}
\end{figure}
\FloatBarrier

\FloatBarrier
\section{Shared dimensionless transport domains}
\label{si:state}
\subsection{Common target and physically paired margins}
The steady cold transport envelope in Fig.~\ref{fig:state}a uses one geometry and target per row: $V_*=2.5\times10^5$ m$^3$ at the bed, $t_w=5$ kyr, $p_{\rm rec}=0.5$ MPa, inherited half-seed 1 km and zero additional tension. Strike $W=100,1000,10000$ m and aperture $b=0.005,0.02,0.1,0.5$ m specify the geometry. Each positive rocky source supplies its dated pressure, gravity, temperature, source enthalpy, actual segregation curve, cavity relaxation curve and initial finite stock $S_0$. Liquid-contact/structure, dry connected melt and positive buoyant segregation are evaluated before assigning the transport state. This leaves Europa and conditional Triton among these self-supplied icy-world references. The cool Enceladus case lacks dry melt; its hot, low-bulk-density reference has no positive buoyant basalt separation. A solid high-pressure ice receiver or a structurally excluded configuration is a separate domain.

For each source, the charged pressure and loading are determined by Eqs.~\ref{eq:u_collect} and \ref{eq:u_open}. The favourable isothermal pressure root $Q_h=Q_s(p_\ell)-Q_a(p_\ell)$ uses main equation (\ref{eq:loaded}). The sealed-source time and extracted stock to charge to $p_{\rm open}$ are
\[
t_{\rm ch}=C_e\int_0^{p_{\rm open}}\frac{\dd p}{Q_s(p)-Q_a(p)},\qquad
V_{\rm ch}=C_e\int_0^{p_{\rm open}}\frac{Q_s(p)\,\dd p}{Q_s(p)-Q_a(p)} .
\]
The integrals use the frozen-date curves and Gaussian quadrature per monotone curve interval; they diverge at the relaxed-pressure boundary. For $Q_h>0$ and attainable opening, use the inventory, wall-exposure and enthalpy terms of main equations (\ref{eq:inventory}) and (\ref{eq:heat}) evaluated at both cumulative bed payloads $v=0,V_*$. The margins used in the steady envelope are
\begin{equation}
\Pi_H=\min\left(\frac{H_{\rm available}(0)}{H_{\rm required}(0)},\
\frac{H_{\rm available}(V_*)}{H_{\rm required}(V_*)}\right),\quad
\Pi_V=\frac{V_{\rm available}}{V_{\rm required}}.
\end{equation}
The minimum connected enthalpy over the path and the initial isothermal rate are favorable to transport. The two prefix tests are necessary for this constant-rate cold-path route; transient flow, evolving source depletion, tip work, side cooling and stress support may tighten its admissibility. No stored elastic burst is added to $Q_h$. If the pressure gate or the loaded-flow gate closes, heat and inventory margins for this loaded trajectory are undefined. $\Pi_P>1$, $\Pi_H\geq1$ and $\Pi_V\geq1$ are the optimistic steady transport criteria. Independently prescribed hot pulses and the paid-contact histories have their own source/time budget below; they do not borrow a thermal margin from a different geometry.

\subsection{Terrestrial shallow-source analog and imposed finite pulse}
The Earth analog is constructed from the same dry source law; it is a controlled shallow melt column rather than an inversion of measured magma supply. The independent Axial seismic distances \citep{yang2024} set $L=1.1,2.3,4.5$ km. Reference density is 3300 kg m$^{-3}$, gravity 9.81 m s$^{-2}$, bed temperature 275 K and bed pressure 18.200775 MPa. $P(z)=P_b+\rho_r gz$; a 3 km connected source interval starts with $\phi_m=0.005$ and increases linearly to 0.20. Temperatures are obtained from the dry Katz solidus/liquidus and melting exponent at local pressure; a linear cold lid connects the top of that interval to the bed. Source parameters use 1, 3, 10 mm grains, 3 km collection radius and 1 km cavity, with coarse-grain 10 and 30 km collection alternatives at the deeper reference lengths and a 0.3 km cavity for the shallow control. The pressure/flux and cavity equations in SI~\ref{si:u_source} apply to the same fields used for wall enthalpy. The 2.3 km, 10 mm, 3 km catchment test at compaction resolutions 384, 768 and 1536 gives unloaded fluxes 0.012286, 0.012283, 0.012282 m$^3$ s$^{-1}$ and $\Pi_H$ of 0.005526, 0.005525, 0.005525 for the 1 km by 0.1 m route; the state classification is preserved. The 54 legacy source/geometry replay points reproduce the earlier unloaded and loaded first-fill heat, opening and hydraulic values to the archived numerical tolerance.

The matched imposed-pulse comparison in Fig.~\ref{fig:state}b uses 5 million m$^3$ at 1550 K (about 30.19 m$^3$ s$^{-1}$ for 46 h) and target $V_*=0.25$ million m$^3$, with $W=10,100,1000$ m and $b=0.1,0.2,0.5$ m. The stipulated held slot and the same dense host 3300 kg m$^{-3}$ provide a separate thermal/geometric comparison, using each selected column's bed pressure and gravity and linear host-temperature profile. The H3 eruption volume and 46 h duration from repeat bathymetry and eruption context \citep{lesaout2020} motivate the magnitude; they are not a measurement of source stock passing through a full cold lid. With no collector accommodation in this prescribed input, $\Pi_V=5\times10^6/(WbL+V_*)$. If $\Pi_V<1$, the required wall exposures at target completion describe the hypothetical full trajectory beyond available time, not delivered material. The Enceladus dense-host control isolates its shorter path even though the adopted porous core does not produce the same buoyant dry source.

In the pressure-only comparison (Fig.~\ref{fig:drivers}c) the 171 km Europa thermal profile, gravity, flux, geometry and total packet remain fixed. The absolute bed pressure is changed from 6.325 to 18.2, 141.2, 177.9 and 280.5 MPa, offsetting the whole pressure profile by the same amount. The intermediate 141.2 MPa value removes the 36.7 MPa ice load at unchanged liquid column in this controlled comparison. The basalt phase interval and pressure-dependent viscosity are evaluated at 1490, 1520 and 1550 K; $Q=30.19$ m$^3$ s$^{-1}$ is externally specified. Thus the near-solidus loss of enthalpy margin arises from the phase interval; the concomitant increase of viscosity also affects the flow resistance, distinct from source buoyancy or imposed gauge driving stress.

\subsection{Recurrence and finite contact margins}
The first finite-strike diffusion rate is $\lambda_{\rm side}=\pi^2\kappa_r/W^2$; $t_{\rm cool}=1/\lambda_{\rm side}$ makes the thermal-memory coordinate in Eq.~\ref{eq:state} explicit. It characterizes side-wall diffusion, not an exact plug-closure time. We evaluate $t_{90}$ by the above charge integral at $p=0.9p_\infty$ using each frozen initial source curve. The actual rest interval up to renewed contact is measured from the coupled history; the wall temperature, liquid fraction and pressure at its end independently establish the state. The terrestrial 4/13 yr reference intervals between the mapped 1998, 2011 and 2015 Axial eruptions \citep{yang2024} are illustrative clocks (same 100 m width), with sealed-source $t_{90}$ calculated from the shallow analog rather than an observed earthquake or eruption recurrence fit.

For the paid branch of Fig.~\ref{fig:state}d each 40 yr contact at its actual date and preceding state has material-equivalent heat debit $V_p$ (SI~\ref{si:recurrence}), fresh fill $V_f$ and finite material available $A_c$ under the midpoint source law and the elastic stored volume. Explicitly, with current available stock $S$, unloaded rate $Q_0$ at contact midpoint, contact duration $t_c$ and initial collector pressure $p_c$,
\[
A_c=S[1-\exp(-Q_0 t_c/S)]+C_ep_c,\quad
\Pi_{V,c}=\frac{A_c}{V_p+V_f},\qquad
\Pi_{H,c}=\frac{[A_c-V_f]_+}{V_p}.
\]
Here the heat margin uses the same sacrificial-material enthalpy conversion as the contact debit. All completed renewed contacts in the four source families, both rest schedules, meet these two gates (minimum ratios about 1.06). The contact conditions supply a heated accessible route prior to actual delivery; the subsequent flow and stock are evolved with their own full mechanical, energy and mass budgets. In the code/data index the self-supplied rows, supplied pulses, absolute-pressure tests, clocks and paid-contact rows remain separate, allowing the same geometry and time support to be recovered for each ratio.

\FloatBarrier

\section{Coupled source, evolving liquid path and receiving boundary}
\label{si:u_path}
\subsection{Advancing wet volume and energy}
The pressure and head budget for wetted length $x$ is equation (\ref{eq:u_head}). Here $B(x)$ is integrated buoyancy in Pa, whereas the local buoyant gradient $(\rho_r-\rho_l)g$ is in Pa m$^{-1}$. The residual term represents additional front/support work or a specified distributed resistance. The receiver head includes both net bed elevation and resistance internal to its growing edifice.
The source pressure is solved implicitly over the accepted interval with the stock and nonnegative-pressure limit in Eq.~\ref{eq:u_collect}. The increment of wet volume is bounded by the next axial cell, the delivery target and the actual available interval. In each wet cell magma amount and enthalpy evolve through axial advection, wall-face heat flux and reversible decompression work; the pressure-dependent phase fraction and viscosity then determine its liquid connectivity. A logarithmic wall-normal mesh exchanges heat with the dated radial background, including the finite-strike lateral-loss scale $\lambda_{\rm side}=\pi^2\kappa_r/W^2$. The update resolves loss of upstream connectivity as an event. Material reaches the bed at the exact full-wet-length condition. Mechanical aperture is evaluated against the local launch and support gates; held-path calculations specify an independent supporting geometry.

The hydraulic equation is Eq.~\ref{eq:u_head}. In a supply-limited, held-path experiment a distributed additional resistance $R_{\rm extra}$ is represented by $b_h=b[R/(R+R_{\rm extra})]^{1/3}$ and the corresponding reduced thermal gap. The resulting gauge-pressure field gives the required compressive-contact fraction $[-\min(p_{\rm gauge}+\sigma_T)]_+/\sigma_C$ and edge demand $[\max(p_{\rm gauge}+\sigma_T)]_+\sqrt{\pi W/2}$. Reference support tests use $\sigma_C=50$ MPa, contact fraction 0.25 and toughness screen 2 MPa$\sqrt{\rm m}$. These comparisons specify the mechanical conditions under which the held geometry is applicable.

\subsection{Bed interface and budgets}
Actual outlet volume and specific enthalpy enter the receiving boundary at the common clock. At reference absolute pressure the finite Robin heat solution is bounded by $c_p$ and $c_p+L_f/\Delta T_{\rm melt}$; the competent depth (1023.15 K) is distinct from solid fraction. The expanding unit must retain its inner liquid core and a Lam\'e shell able to carry the load. Internal-pipe heat and resistance determine the head $p_{\rm rec}$, including net elevation in Eq.~\ref{eq:u_head}. The conservative placement kernel integrates to one; it transfers actual retained material to a growing mapped edifice. Cold/mixed water evolves through $C_w=\rho_wc_{p,w}V_w$ and renewal loss $E_w/\tau_w$. The interface records both retained and unsupported receiver-phase material; the total delivered volume and enthalpy enter the budget irrespective of the deposition classification.

The transferred-volume identity is equation (\ref{eq:ledger}). Route energy accounts for liquid/solid magma, walls, wall loss, reversible pressure work, preparation residual enthalpy, old plugs and bed export. Source enthalpy includes thermal conditioning of its stored liquid at the evolving host temperature. Integrated collector and gravitational work balance hydraulic losses, kinetic change, front traction and receiving work; the work dissipation is separately recorded. These balances couple the physical submodels through the actual exchanged rates and enthalpies.

\begin{table}[H]\centering\small
\caption{Transferred quantities and their roles in the coupled calculation.}
\label{tab:interfaces}
\begin{tabularx}{\textwidth}{@{}l l X@{}}\toprule
Symbol&Unit&Physical role\\\midrule
$p_c,P_*,p_\infty$&Pa&Rock-lithostatic excess, stopping and relaxed collector pressure.\\
$P_b,p_{\rm rec}$&Pa&Absolute ocean pressure and gauge head required at the bed.\\
$S$; $Q_s,Q_a,Q_f$&m$^3$; m$^3$ s$^{-1}$&Finite stock; segregation, accommodation and route rates.\\
$W,b,b_h$&m&Strike, mechanical and equivalent hydraulic aperture.\\
$B(x),h_l,f_l$&Pa, J kg$^{-1}$, --&Integrated buoyancy, magma enthalpy and liquid fraction.\\
$V_{\rm wet},V_{\rm plug},V_p$&m$^3$&Current route magma, earlier frozen material and sacrificed preparation material.\\
$V_b,V_r$; $H_g-w$&m$^3$; m&Bed arrival and retained material; net relief.\\\bottomrule
\end{tabularx}\end{table}
\FloatBarrier

\section{Thermal resetting and repeated-delivery resource budget}
\label{si:recurrence}
The repeated material histories at the shared 0.1 Myr endpoint using the Gaussian receiver are presented in main Table~\ref{tab:reset} and Fig.~\ref{fig:repeat}. Every paid-access 5 kyr case has 20 positive arrival episodes; every 10 kyr paid-access case has 10. Both sealed and cold-adjacent branches share the initial preparation and the first opportunity's delivery. Table~\ref{tab:paidmap} separately reports the mapped receiver results.
\begin{figure}[!htbp]\centering\includegraphics[width=.98\textwidth]{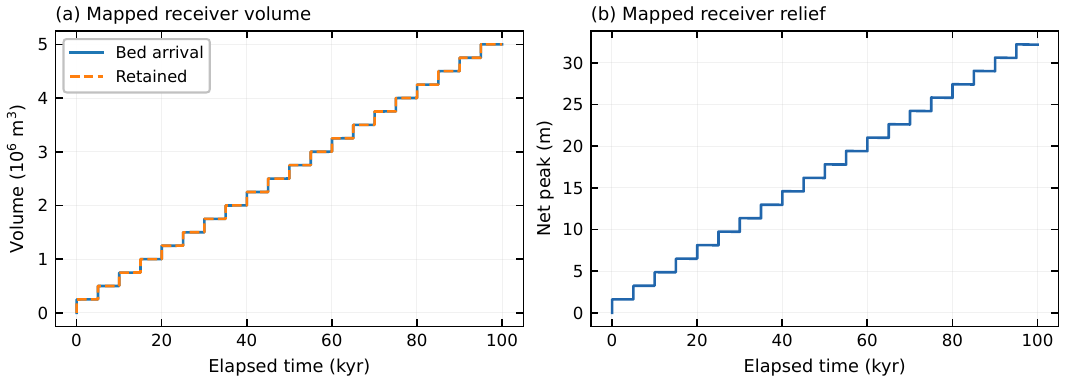}
\caption{Distinct direct central-mapped hot Europa 100 km paid-access history, including the final dormancy at 0.1 Myr. (a) Actual bed arrival and retained material. (b) Net peak height on the same run, as the receiver evolves. These are mapped receiver data; the corresponding purely Gaussian controls and budgets are given in main Fig.~\ref{fig:repeat} and Table~\ref{tab:reset}.}
\label{fig:u_cycle}\end{figure}
The common rest update continues sealed source recharge, solid plug cooling, wall diffusion and receiver heat/foundation evolution. A previously accessible contact maintained for $t_p$ has target wall profile
\[
T_p(x,z)=T_{\rm bg}(z)+[T_{\rm in}-T_{\rm bg}(z)]_+
\operatorname{erfc}\left(\frac{x}{2\sqrt{\kappa_rt_p}}\right).
\]
The incremental heat above the existing wall plus lateral-loss estimate gives $E_p$, and $V_p=E_p/[\rho_l(h_{\rm in}-c_p\bar T_{\rm bg})]$ debits the finite hot stock. Refill costs $WbL$. Both must be supplied inside the 40 yr contact interval, allowing for elastic storage, the contemporaneous pore rate, accommodation and the declining stock. The initial and repeated preparation intervals are included in elapsed time. The mapped histories use the same source-law and thermal path as the paid-access branch, with local previously resolved mass-normalized raster and head-selected receiver. Sealed, cold-adjacent and paid-access experiments distinguish persistence of the magma source from persistence of the liquid path.

\begin{table}[H]\centering\small
\caption{Finite source and receiver budget for the mapped central-direct 5 kyr histories. Available stock/heat preparation, prior plug volume and bed emission have distinct ledgers. The final peak includes the last 5 kyr-scale rest period.}
\label{tab:paidmap}\begin{tabularx}{\textwidth}{@{}Xrrrrrr@{}}\toprule
Source / catchment&$Q_0$ (m$^3$ s$^{-1}$)&Stock (km$^3$)&Prep. (km$^3$)&Plug ($10^6$ m$^3$)&Bed ($10^6$ m$^3$)&Peak (m)\\\midrule
\UCycleRows
\bottomrule\end{tabularx}\end{table}

The 60 km and 100 km focused geometries have initial stocks determined by the same connected interval and 2 percent accessible fraction. The mechanical collector radius is 3 km, whereas the catchment areas are $\pi(60\ \mathrm{km})^2$ and $\pi(100\ \mathrm{km})^2$. The 100 m by 0.2 m path's refill volume is separately debited from the preheat budget. The central Europa full-target mapped history at 100 kyr retains 5.000 million m$^3$ and $32.16$ m net peak; its upper wall is 278.76 K with $f_l=0$. The Gaussian receiver uses a different normalised footprint and supplies the separate sealed / cold-adjacent / paid 5 and 10 kyr comparison in main Table~\ref{tab:reset}. In the direct mapped histories the same consumed source enthalpy is used for preparation and bed export. The exact receiver head changes with its growing relief (Fig.~\ref{fig:u_cycle}).

\FloatBarrier
\section{Independent source and fracture reconstructions}
\label{si:reference_controls}
\subsection{Different thermal and porous-flow closure}
The independent 108-design source set is reconstructed from 29 earlier dated thermal activation states of 85--650 km rock-radius bodies ($\rho_r=3300$ kg m$^{-3}$). The thermal enthalpy equation follows Eq.~\ref{eq:u_rock} with cold assembly at 1.45 Myr after CAIs, $^{26}$Al power $2.13\times10^{-7}$ W kg$^{-1}$ at CAI time, 0.717 Myr half-life, reference long-lived power $5.7925\times10^{-12}$ W kg$^{-1}$ and layered tidal response \citep{beuthe2015,katz2003}. Assigned water mass fractions 0.015, 0.032 and 0.060, grain diameters 75/125/250 $\mu$m, $\phi\leq0.225$, $\mu_l=10$ Pa s, collector radius 3 km and 2 percent accessible melt define the cases. These yield actual launch flux, pressures, thermal profiles and material inventories for a separate geometry and permeability law.
\[
k_p=d_g^2\phi^{2.6}/58,\quad K=k_p/\mu_l,\quad C_m=\eta_m(1/\phi+4/3),
\quad q/K-\partial_z(C_m\partial_zq)=(1-\phi)\Delta\rho_mg.
\]
The outlet uses its prescribed local pressure traction, the bottom zero relative flux, and the material flux $(1-\phi)q$ \citep{miller2014,mckenzie1984}. Finite radial cavity accommodation has $R_v=4\eta_c/(3V_c)$ and $C=14.137$ m$^3$ Pa$^{-1}$. At the homogeneous limit, $D=(1-\phi)\Delta\rho_mg$, $\ell_c^2=KC_m$, $u=L/\ell_c$ and
\[
Q_0=(1-\phi)A_c KD(1-\operatorname{sech}u),\quad
P_s=D\ell_c\tanh(u/2),\quad
p_\infty=\frac{Q_0}{Q_0/P_s+1/R_v}.
\]
At fixed material properties $Q_0,P_s\propto g$; opening requires adequate stock $S\geq Cp_{\rm req}$ and time to reach $p_{\rm req}$. For the 15 km matched column, $\phi=0.03$, $d_g=125\ \mu$m, $\ell_c=32.02$ m, $50$ m radius flaw and tension 50 kPa give $p_{\rm req}=75.33$ kPa. In its non-relaxing limit Earth reaches the threshold in 121 yr, $g_{\rm crit}=4.85$ m s$^{-2}$; the reference strongly accommodating case increases the threshold. Table~\ref{tab:matchedcharge} compares relaxed/elastic charge and supply in the three matched environments. Altering flaw radius, toughness, tensile stress or mobility shifts the boundary.

\begin{table}[!htbp]\centering\small
\caption{Independent homogeneous 15 km source at matched material properties, 50 m access radius and favorable 50 kPa resolved tension. Values at 10 kyr. R and N denote relaxing and non-relaxing collectors: $p_c^{\rm N}=P_s$, $p_{\rm req}=75.331$ kPa and required elastic/fill stock $1.065\times10^6$ m$^3$; accessible stock is $2.262\times10^9$ m$^3$. The Europa control uses the earlier shallow-reference mechanical gravity (1.384 m s$^{-2}$).}
\label{tab:matchedcharge}
\begin{tabular}{@{}lrrrrrrr@{}}
\toprule
Reference & \shortstack{$g$\\(m s$^{-2}$)} & \shortstack{$Q_0$\\($10^{-5}$ m$^3$ s$^{-1}$)} & \shortstack{$P_s$\\(kPa)} & \shortstack{$p_c^{\rm R}$\\(Pa)} & \shortstack{$V_{\rm el}^{\rm R}$\\(m$^3$)} & \shortstack{$V_{\rm el}^{\rm N}$\\($10^6$ m$^3$)} & $P_s/p_{\rm req}$\\\midrule
Earth & 9.810 & 38.591 & 152.34 & 1.752 & 24.8 & 2.1537 & 2.022\\
Europa & 1.384 & 5.446 & 21.50 & 0.2473 & 3.5 & 0.3039 & 0.2854\\
Enceladus & 0.128 & 0.504 & 1.99 & 0.02287 & 0.323 & 0.0281 & 0.02639\\
\bottomrule
\end{tabular}
\end{table}

\begin{figure}[!htbp]\centering\includegraphics[width=.97\textwidth,height=.59\textheight,keepaspectratio]{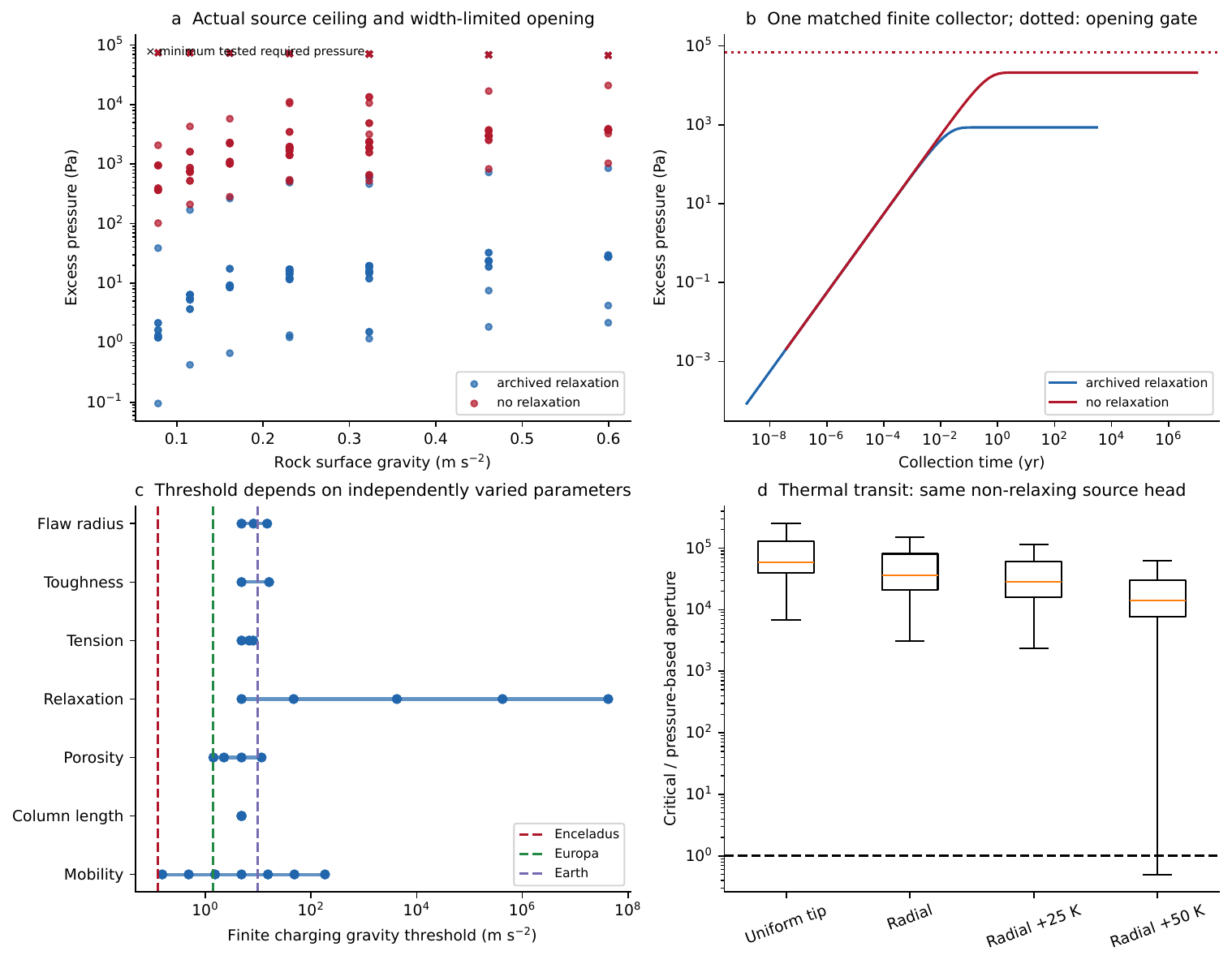}
\caption{Finite reconstructed source pressure versus geometric opening need, actual charging histories, matched finite gravity thresholds and independent thermal-transit ratios. The favorable pre-accessible disk is distinguished from dry or width-limited opening.}\label{fig:sourcechain}
\end{figure}

Across the 108 actual source identities, the maximum pressures are 0.855 kPa with relaxation and 20.99 kPa without. Each source contributes twelve width-limited or intact-wall comparisons (four tension/strength, eight toughness/tension/buoyancy); all are below the respective opening condition, 0/1296 for each accommodation law. As geometry and tensile state change, liquid-accessible disks may be filled under a much smaller pressure demand. Figure~\ref{fig:sourcechain} retains the full reconstructed source-opening and transit comparison.

\subsection{Finite accessible disk and detached ascent}
\begin{figure}[!htbp]\centering\includegraphics[width=\textwidth]{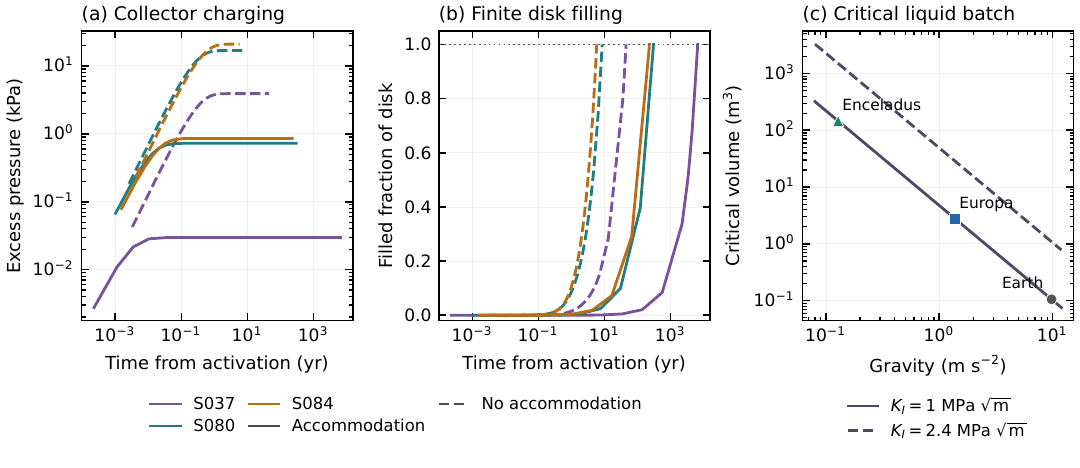}
\caption{Independent finite sources, accessible disks and the critical detached volume. (a) Actual collector excess pressure and (b) the filled fraction of each disk versus the time from source activation, with and without viscous accommodation. (c) Analytical critical liquid volume as a function of gravity for a density contrast of 500 kg m$^{-3}$, plane-strain modulus of 80 GPa and the two indicated toughness values. Table~\ref{tab:disk} gives the detached cold-front fractions and insulated reference.}\label{fig:connected}
\end{figure}

\begin{figure}[!htbp]\centering\includegraphics[width=.97\textwidth,height=.58\textheight,keepaspectratio]{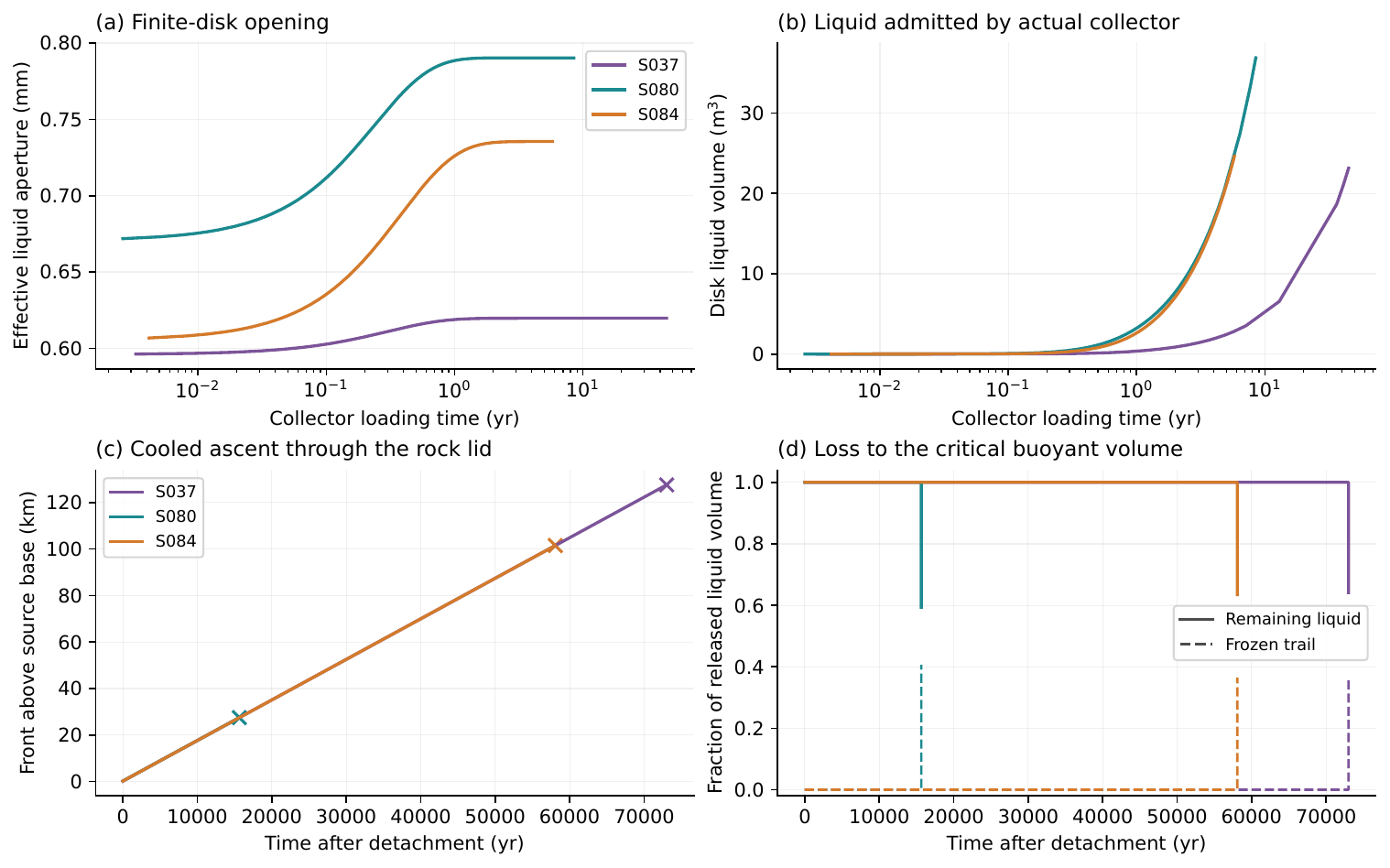}
\caption{Causal disk loading and detached ascent from its actual liquid volume. Crosses show the location where the remaining liquid falls to its critical buoyant volume.}\label{fig:trailtime}
\end{figure}

For a buoyant liquid disk in an elastic host, the reduced reference critical geometry of \citet{davis2020,davis2023} uses
\[
a_b=\left(\frac{3\sqrt\pi K_I}{8B}\right)^{2/3},\quad
p_b=\frac{2Ba_b}{3},\quad
V_{\rm crit}=\frac{32Ba_b^4}{9E'}.
\]
This gives $V_{\rm crit}\propto g^{-5/3}$ and the early finite-batch speed proportional to $g^2V$ at fixed properties (Fig.~\ref{fig:connected}c). The connected disk solution couples source charging and filling through the source law of this section, pressure-root step and actual liquid aperture $b_l=f_lb_{\max}$. The prescribed favorable disk uses independently resolved 50 kPa tension. Normal transient conduction uses the actual radial host temperature; separate detached ascent uses the melted liquid volume in the disk. Warming of each new wall surface debits sensible/latent magma heat. The three reconstructed sources fill accessible disks on different timescales (Fig.~\ref{fig:connected}a,b). Their cooled detached batches advance between 0.326 and 0.750 of the rock lid, relative to the full-traversal insulated reference (Table~\ref{tab:disk}). The liquid aperture, frozen trail and residual mobile volume are tracked in Fig.~\ref{fig:trailtime}.

\begin{table}[H]\centering\small
\caption{Reconstructed finite disks and detached-batch ascent. R and N denote the relaxing and non-relaxing sources; $a$ and $V_d$ are the filled disk radius and liquid volume. The cold front fractions use 512 ascent segments; the insulated reference is the full-lid traversal limit (unity).}
\label{tab:disk}\setlength{\tabcolsep}{4pt}\begin{tabular}{@{}lrrrrrrr@{}}\toprule
 & & & & \multicolumn{2}{c}{Fill time (yr)} & \multicolumn{2}{c}{Front/lid}\\
\cmidrule(lr){5-6}\cmidrule(l){7-8}
Source&$a$ (m)&$V_d$ (m$^3$)&$p_{\max}$ R/N (kPa)&R&N&Cold&Insulated\\\midrule
\ADiskRows
\bottomrule
\end{tabular}\end{table}

\FloatBarrier

\subsection{Continuous feed and wall memory in the same independent sources}
The comparison in Table~\ref{tab:cold_transport_geometries} uses the same three relaxing source states to separate the effects of continuous supply and aperture. Sustained inflow through the disk-mean aperture advances less far than the corresponding detached finite batch. A uniform disk-center upper aperture extends penetration, while the independently held 0.50 m by 1 m cold slot advances through a larger fraction of the lid. The wider held slot is an elastic-support endmember rather than the disk's actual aperture. The ongoing rate follows the same reconstructed $Q_0,P_s,C,R_v$ law with its actual disk and finite accessible stock. In the common source--slit model the same three relaxing launch states are imported from their resolved histories with the original radial wall temperatures interpolated over 513 depths and the common pressure-dependent crystallization and receiver closure. They reach intermediate fractions of the cold lid before liquid arrest (Fig.~\ref{fig:u_thermal}).

\begin{table}[H]\centering\small
\caption{Cold-path penetration of the three relaxing finite sources, comparing detached ascent with continuous feed in three slit geometries. Uniform slit apertures use either the disk-mean aperture or the disk-center upper bound. The independent held slit has 0.50 m aperture and 1 m strike. All front/lid values are dimensionless maximum advance fractions of the corresponding full rock lid.}
\label{tab:cold_transport_geometries}
\begin{tabular}{@{}lrrrrrr@{}}\toprule
 & Detached & \multicolumn{2}{c}{Mean-aperture slit} & \multicolumn{2}{c}{Center-aperture slit} & Held slit\\
\cmidrule(lr){3-4}\cmidrule(lr){5-6}
Source & Front/lid & $b$ (mm) & Front/lid & $b$ (mm) & Front/lid & Front/lid\\\midrule
\AColdGeometryRows
\bottomrule\end{tabular}
\end{table}

\begin{figure}[!htbp]\centering\includegraphics[width=\textwidth]{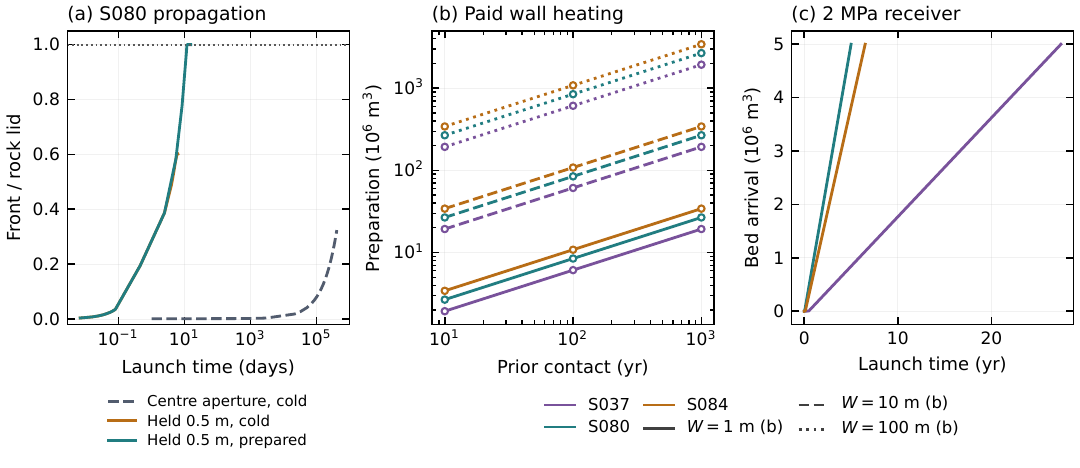}
\caption{Continuous feed from the independently reconstructed finite sources. (a) S080 front position at positive launch times in the cold uniform center-aperture slit and the independently held 0.50 m by 1 m slit, with and without prior paid heating. (b) Minimum magma-equivalent preparation inventory versus contact duration for each source, at 0.50 m aperture and the indicated strike widths. (c) Actual conditional bed arrival through the held 0.50 m by 1 m prepared slit into a 2 MPa receiver after prior 100 yr thermal contact; elapsed launch time excludes preparation. Table~\ref{tab:cold_transport_geometries} compares all three cold-source geometries with detached ascent.}\label{fig:advancing}
\end{figure}

Prior accessible heat changes the independently held path (Fig.~\ref{fig:advancing}a). At 0.50 m aperture and 10 m strike, a 100 yr contact demands $8.41\times10^{16}$, $1.14\times10^{17}$ and $1.50\times10^{17}$ J for the three sources--61.1/84.5/108.2 million m$^3$ of fully sacrificed magma (Fig.~\ref{fig:advancing}b). Source-rate bounds require at least 329/100/139 yr, respectively. The warmer narrow 0.50 m by 1 m pathway with prior 100 yr contact sustains conditional delivery even against a 2 MPa bed head (Fig.~\ref{fig:advancing}c); the largest sustained rate is $0.03193$ m$^3$ s$^{-1}$, compared with 30.19 m$^3$ s$^{-1}$ for five million m$^3$ in 46 h. Sensible-only preheating requires considerably more volume and time. These results independently identify the heat/time/source constraints of the accessible warm-path regime.

The pressure--volume atlas evaluates accessible versus dry flaws (10--3000 m radius, both relaxation limits, two toughness values and tension from $-50$ to $100$ kPa). Neutral-tension dry passes in two extremely favorable 3 km flaws still require hundreds of thousands of m$^3$ of connected liquid. An S084 200 m disk with 50 kPa tension requires much less pressure for accessible filling but has an upper center-aperture estimate of 0.635 mm. Finite held-slot heat tests at the S084 rate require a warm host near the crystallization threshold. This separates the role of pressure reversals from heat and connectivity as in the common model.

\begin{figure}[!htbp]\centering\includegraphics[width=.96\textwidth,height=.61\textheight,keepaspectratio]{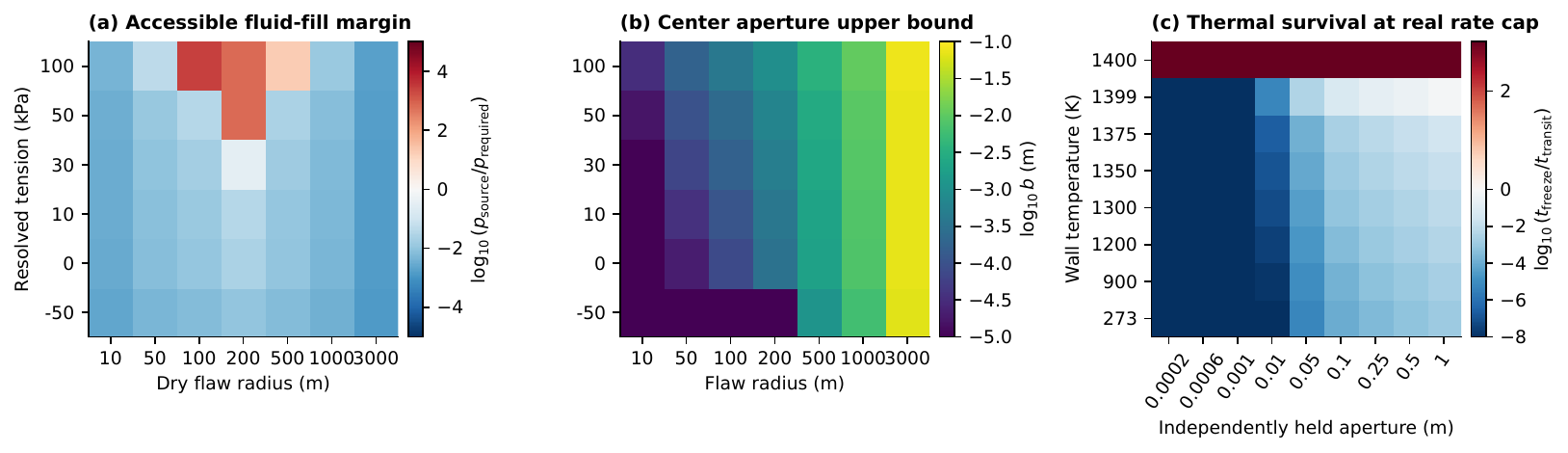}
\caption{Independent S084 pressure/volume and geometry screens and full-contact thermal-transit margin. The favorable tensile-accessibility and aperture conditions precede cold/warm path survival.}\label{fig:jointgate}
\end{figure}

\begin{figure}[!htbp]\centering\includegraphics[width=.93\textwidth,height=.56\textheight,keepaspectratio]{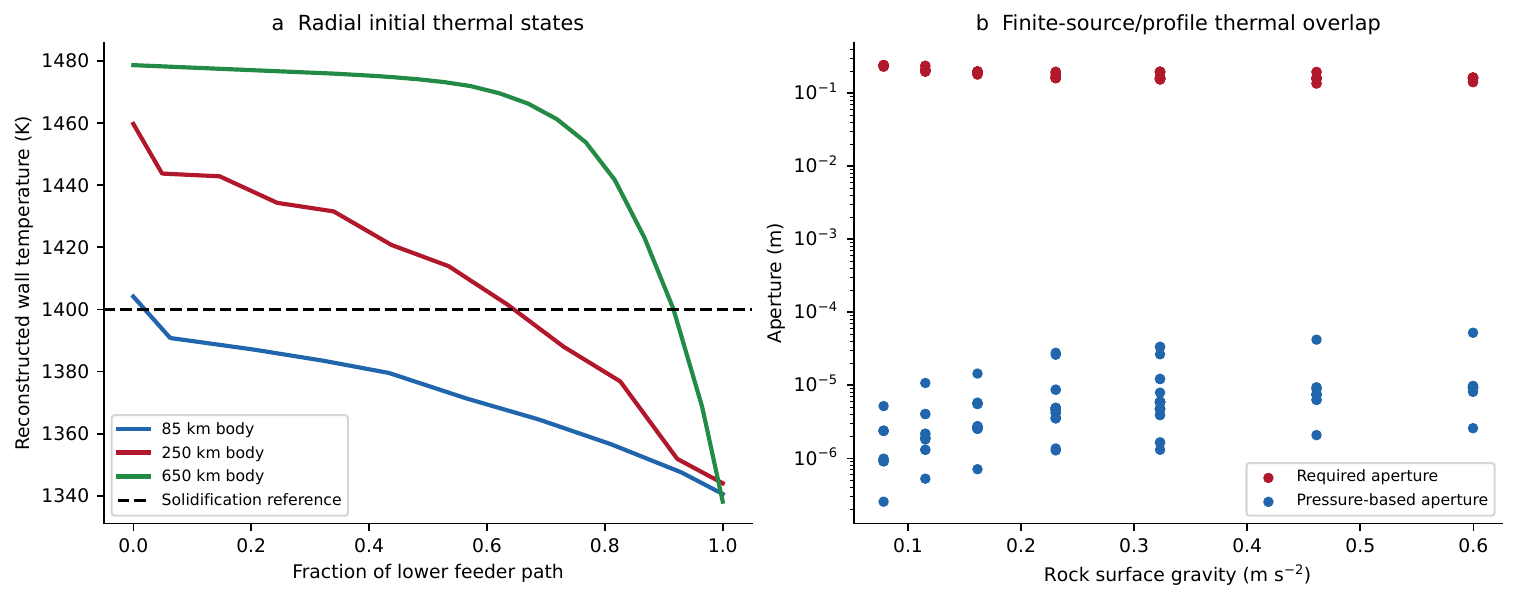}
\caption{Independent radial wall-temperature and finite-source transit comparisons. Depth uses the actual path length of each host profile. The favorable pressure-based aperture is compared with the conductive-transit requirement.}\label{fig:radialwall}
\end{figure}

An independently initially filled 100 m wide radial slot driven by these source pressures closes in about 1.1--1.2, 6.7--7.6 and 26.8--29.9 days as aperture increases from 0.10 to 0.25 to 0.50 m. A maintained-pressure 2 MPa transient control retains heat between 90 day pulses and 180 day rest intervals and shows a later closure than the reset-wall case (S084: 278.8 versus 270.0 days). Those outlet volumes are lower-route transfers. Their progression complements the first-fill and whole-lid repeated-access results.

\begin{figure}[!htbp]\centering\includegraphics[width=.96\textwidth,height=.59\textheight,keepaspectratio]{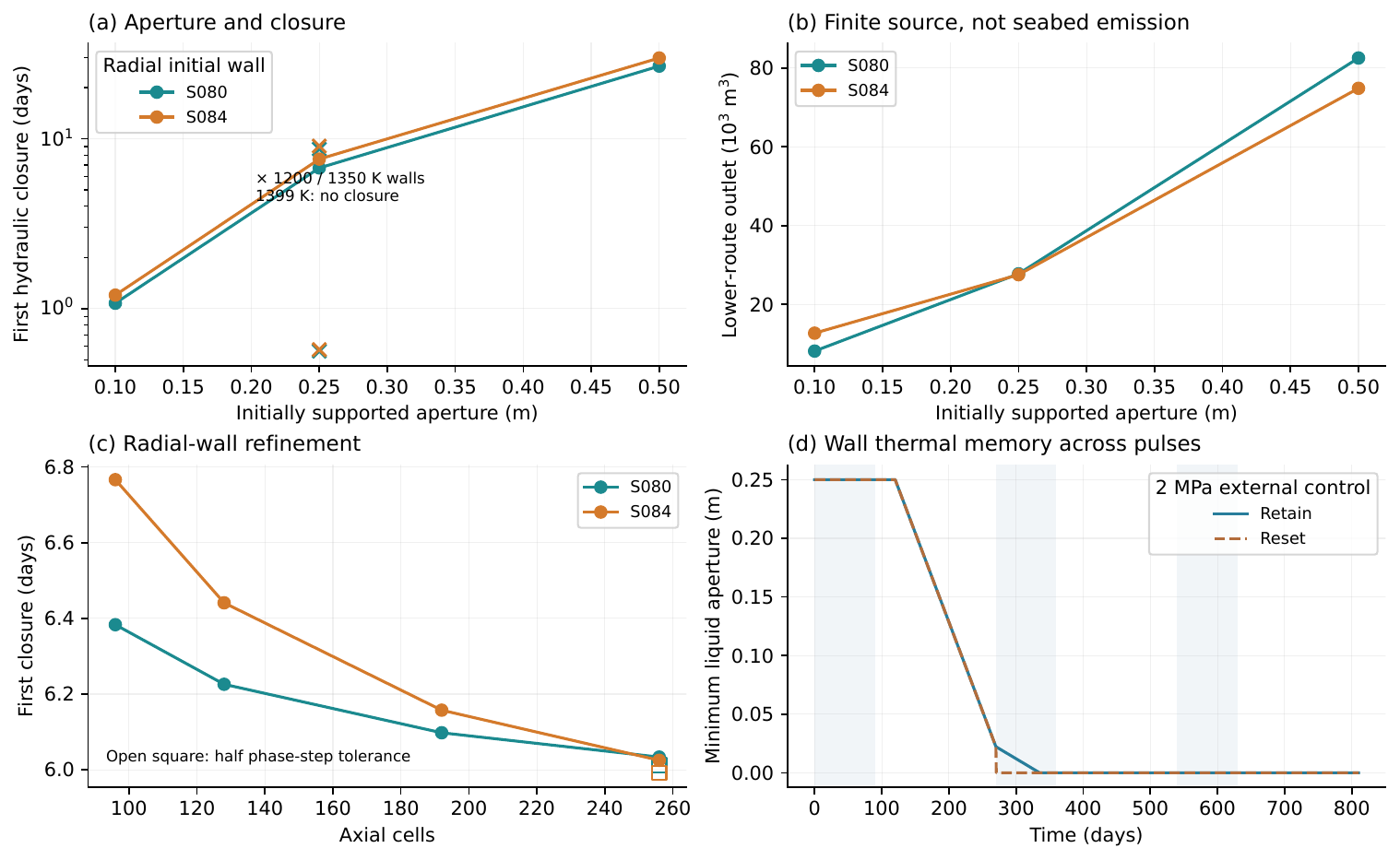}
\caption{Transient radial-path wall heating and finite liquid closure. Aperture/time and temperature sensitivities use reconstructed source states, whereas the repeated-pulse wall-memory result uses the separate maintained-head control.}\label{fig:transientwall}
\end{figure}
\FloatBarrier

\section{Numerical verification and physical uncertainty}
\label{si:u_numerics}
\subsection{Analytic and grid limits of the common model}
In the homogeneous diffusion-only compaction limit, stopping pressure is $B\delta_c\tanh[L/(2\delta_c)]$ and free rate $AB(k_\phi/\mu_l)[1-\operatorname{sech}(L/\delta_c)]$, $\delta_c^2=\zeta k_\phi/\mu_l$. At 512 cells the independent nonlinear implementation gives relative errors of $6.00\times10^{-6}$ in pressure and $6.29\times10^{-7}$ in flux (rounded). Newtonian cavity and Bingham pipe limits are also verified; finest wall-normal Robin heat integral at one year agrees within 0.223 percent with the analytic result and its relative discrete energy residual is below $10^{-10}$. Dated rock temperatures and nonlinear source flux are separately refined to 6144 thermal cells and 1536 compaction cells. The 136 primary, 35 mapped and 40 refinement coupled calculations have maximum relative material and thermal residuals below $6\times10^{-13}$ and $2.0\times10^{-8}$; source-to-path and receiving heat transfers close within $3\times10^{-15}$, and integrated pressure work within $5\times10^{-12}$. The residuals measure discrete conservation; supported endpoint distances and temperatures measure solution sensitivity.

\begin{table}[H]\centering\small
\caption{Axial/wall refinement of cold held slots. Front length is the maximum fraction of the original lid. None of the four representative families reaches the bed in the specified time window. At 1024 cells the fine-grained Europa run ends at the 1000 yr time window; the other three end at the liquid-connectivity threshold (main Fig.~\ref{fig:front}).}
\label{tab:u_front_ref}\begin{tabularx}{\textwidth}{@{}Xrrr@{}}\toprule
Source&Axial cells&Front (\%)&Bed arrival (m$^3$)\\\midrule
\UFrontRefineRows
\bottomrule\end{tabularx}\end{table}

The repeated prepared Europa 60 km and Triton 60 km 128/256 cell pairs differ in arriving volume by less than 0.01 percent. At matched 5 million m$^3$ retained volume and all twenty pulses, the three central mapped grid levels 64/128/256 give peaks 31.77/32.28/32.48 m in the representative refined case. Exact vanishing-increment timing can shift with the phase and support boundary. The final mapped peak is distinguished from the Gaussian placement comparison.

\subsection{Independent disk, advancing-flow and thermal tests}
For the independent coupled disk, conservative advection uses a characteristic-bound Rusanov flux (the short-wave relative flux speed approaches $-q_0$); nonnegative porosity/volume and $\phi\leq0.225$ bound the update. Pressure and temperature roots, explicit exposed wall fractions and 16/24/32 normal-cell conduction ensure the finite material and energy transfer. Across its 32 primary source/geometry comparisons, maximum global relative volume, heat and pressure-work errors are $1.25\times10^{-11}$, $8.1\times10^{-10}$ and $2.3\times10^{-11}$. In incompletely filled, large nearly equilibrated disks, the incremental net-heat residual can be as large as 1.71 of a nearly cancelling heat increment (absolute discrepancy up to 186 J; state-scaled discrepancy $6.0\times10^{-8}$). The filled-disk/thermal stopping classification is stable through joint refinement, while the small net-heat increment in those large-disk endpoints has weaker local precision.

\begin{table}[H]\centering\small
\caption{Independent disk convergence: terminal thermal advance at 512 axial segments. These cases finish accessible disk filling and reach the cooled critical liquid-volume stop.}
\label{tab:disk_refinement}
\begin{tabular}{@{}llrrrr@{}}\toprule
Source&Accommodation&Column/wall cells&Fill (yr)&Front (km)&Bed (m$^3$)\\\midrule
S080&None&72/16&8.4874&27.503&0\\
S080&None&144/24&8.4809&27.174&0\\
S080&None&288/32&8.4786&27.174&0\\
S084&None&72/16&5.7842&101.476&0\\
S084&None&144/24&5.7848&101.134&0\\
S084&None&288/32&5.7851&101.134&0\\
S084&Viscous&72/16&227.681&101.476&0\\
S084&Viscous&144/24&227.722&101.134&0\\
S084&Viscous&288/32&227.803&101.134&0\\\bottomrule
\end{tabular}\end{table}

The independent advancing volume comparisons refine both axial sampling and permitted time/step size. Conservative wetted-area integration retains the square-root exposure-age heat increments, pressure and finite stock. Source laws, physical widths, host temperatures and heat closure are kept fixed across grid levels. The cold arrest/warm conditional-arrival ordering survives these tests. The common model uses its separate pressure-dependent latent interval; the earlier independently heated disk uses a phase reference at 1400 K. The joint and independent evidence therefore tests the same mechanism with differing thermal and mechanical constitutive assumptions.
\FloatBarrier

\section{Independent published data and physical context}
\label{si:external}
Europa dike propagation calculations by \citet{bland2022} establish combinations of injection and lithospheric structure needed for ascending magma; \citet{green2025} determine source and dike transport with independent thermal histories. The twelve archived Green model trajectories \citep{green2024data} record the maximum fraction of each lithospheric route traversed and the associated simulation interval (Table~\ref{tab:green}). Their partial penetrations support the separation of melting and successful fracture transport. The model depth, rate and geometry choices differ from our present coupled source. The Axial seismic reservoir roof 1.1--2.3 km and deeper lens depth to 4.5 km \citep{yang2024} independently constrain the terrestrial distance used in Fig.~\ref{fig:u_thermal}a. Surface and interior stress constraints \citep{byrne2026} inform plausible resolved tensions rather than specifying a deep accessible trace. The mapped fissures \citep{axialoutlines,lesaout2020} independently constrain an eruptive strike scale; they enter geometric comparisons, not an assumed satellite fracture width. These external anchors constrain distinct stages of the common material/heat mechanism.

\begin{table}[!htbp]\centering\footnotesize
\setlength{\tabcolsep}{4pt}
\caption{Independent published Europa trajectories \citep{green2025,green2024data}. Maximum recorded penetration is normalized by each trajectory's own lithospheric path length; duration is the recorded interval. All twelve records remain below complete transit.}
\label{tab:green}
\begin{tabular}{@{}lrr@{\hspace{6mm}}lrr@{}}
\toprule
Case & Max. fraction & Duration (Myr) & Case & Max. fraction & Duration (Myr)\\\midrule
LL1280MaNR & $7.18\times10^{-6}$ & 179.8 & Perm4 & $2.34\times10^{-2}$ & 153.8\\
LL3GaNR & $1.23\times10^{-2}$ & 112.4 & Perm5 & $7.48\times10^{-2}$ & 98.3\\
LL500MaNR & $1.16\times10^{-5}$ & 206.9 & Perm6 & $1.44\times10^{-2}$ & 200.5\\
Perm1Lo & $8.98\times10^{-4}$ & 112.5 & Reference1NR & $9.10\times10^{-3}$ & 217.9\\
Perm2 & $1.47\times10^{-5}$ & 197.9 & mEta100NR & $6.30\times10^{-7}$ & 210.1\\
Perm3 & $1.58\times10^{-2}$ & 160.0 & mEta1NRLoYield & $1.85\times10^{-5}$ & 152.4\\
\bottomrule
\end{tabular}
\end{table}

\FloatBarrier
\end{document}